%% file: kinetic_gvd_arxiv.tex
\documentclass[letter,11pt]{journal}
\usepackage{xspace}
\usepackage{algorithmic}
\usepackage{amsmath}
\usepackage{booktabs}
\usepackage{enumitem}
\usepackage[colorlinks]{hyperref}
\usepackage{graphicx, wrapfig}
\usepackage{amssymb,amsthm,amsmath}
\usepackage[dvipsnames]{xcolor}

\usepackage[left=1in,top=1in,right=1in,bottom=1in]{geometry}
\usepackage{microtype}

\hypersetup{allcolors = black}

\graphicspath{ {./figures/} }

\newcommand{\myremark}[4]{\textcolor{blue}{\textsc{#1 #2:}} \textcolor{#4}{\textsf{#3}}}
\renewcommand{\myremark}[4]{}

\newtheorem{theorem} {Theorem}
\newtheorem{lemma}[theorem] {Lemma}
\newtheorem{corollary}[theorem] {Corollary}

\newcommand{\thmheadfont}{\bfseries}
\newenvironment{repeatenv}[2]%
  {\smallskip\noindent {\thmheadfont #1~\ref{#2}.}\ \slshape}
  {\normalfont}

\renewcommand{\subparagraph}[1]{\smallskip\noindent\textbf{#1}}

\newcommand{\etal}{et al.\xspace}
\newcommand{\eps}{\ensuremath{\varepsilon}\xspace}

\newcommand{\mkmcal}[1]{\ensuremath{\mathcal{#1}}\xspace}

\renewcommand{\S}{\mkmcal{S}}

\newcommand{\Lst}{\mkmcal{L}}
\newcommand{\V}{\mkmcal{V}}

\newcommand{\mkmbb}[1]{\ensuremath{\mathbb{#1}}\xspace}
\newcommand{\R}{\mkmbb{R}}

\newcommand{\Time}{\mkmbb{T}}
\newcommand{\geod}{\pi\xspace}
\newcommand{\geodlen}{\pi\xspace}
\newcommand{\dist}{\geod}

\DeclareMathOperator{\VD}{VD}
\DeclareMathOperator{\SPM}{SPM}
\DeclareMathOperator{\ep}{ep} 

\title{Kinetic Geodesic Voronoi Diagrams\protect\\ in a Simple Polygon}

\author{Matias Korman\thanks{Siemens Electronic Design Automation, Wilsonville, Oregon, USA 
  \texttt{matias.korman@siemens.com}
  }
  \and
  Andr\'e van Renssen\thanks{
    School of Computer Science, University of Sydney, Sydney,
    Australia \texttt{andre.vanrenssen@sydney.edu.au}
  }
  \and
  Marcel Roeloffzen\thanks{Dept. of Mathematics and Computer Science, TU Eindhoven,
    Eindhoven, The Netherlands \texttt{m.j.m.roeloffzen@tue.nl}
  }
  \and
  Frank Staals\thanks{Dept. of Information and Computing Sciences, Utrecht
    University, Utrecht, The Netherlands \texttt{f.staals@uu.nl}
  }
}

\date{}

\begin{document}
\maketitle

\begin{abstract}
  We study the geodesic Voronoi diagram of a set $S$ of $n$ linearly
  moving sites inside a static simple polygon $P$ with $m$
  vertices. We identify all events where the structure of the Voronoi
  diagram changes, bound the number of such events, and then develop a
  kinetic data structure (KDS) that maintains the geodesic Voronoi
  diagram as the sites move. To this end, we first analyze how often a
  single bisector, defined by two sites, or a single Voronoi center,
  defined by three sites, can change. For both these structures we
  prove that the number of such changes is at most $O(m^3)$, and that
  this is tight in the worst case. Moreover, we develop compact,
  responsive, local, and efficient kinetic data structures for both
  structures. Our data structures use linear space and process a
  worst-case optimal number of events. Our bisector and Voronoi center
  kinetic data structures handle each event in $O(\log^2 m)$
  time. Both structures can be extended to efficiently support
  updating the movement of the sites as well. Using these data
  structures as building blocks we obtain a compact KDS for
  maintaining the full geodesic Voronoi diagram.
\end{abstract}

\thispagestyle{empty}
\clearpage
\setcounter{page}{1}

\section{Introduction}

Polygons are one of the most fundamental objects in computational geometry. As
such, they have been used for many different purposes in different
contexts. Within the path planning community, polygons are often used to model
different regions. A simple example is planning the motion of a robot in a building:
 we can model all possible locations that a robot can reach
by a polygon (the walls or other obstacles would define its boundary).
Then, the goal is to find a path that connects the source point with its
the destination and that minimizes some objective function. There are
countlessly many results that depend on the exact function used (distance
traveled~\cite{guibas1987spm}, time needed to reach a destination~\cite{m-spaop-93}, number
of required turns~\cite{suri}, etc.) 

Paths that
minimize distance are often called \emph{geodesics}. In this paper, we consider one of the most natural metrics: the domain is a simple polygon $P$ and paths are constrained to stay within the closure of $P$. Under this setting it is well known that given two points $s,t \in P$ there exists a unique geodesic $\geod(s,t)$ connecting the two points. Moreover, $\geod(s,t)$ is a simple polygonal chain whose vertices (other than the first and last) are reflex vertices of $P$. Thus, we define the {\em geodesic distance} function $\geodlen{s,t}$ between $s$ and $t$ as the the sum of Euclidean lengths of the segments in $\geod(s,t)$. With the properties mentioned above, it follows that the geodesic distance is a proper distance and is well defined~\cite{m-spaop-93}.

 \begin{figure}[tbh]
   \centering
   \includegraphics{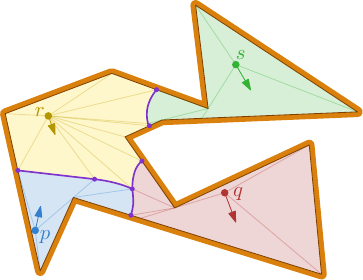}
   \caption{The (augmented) geodesic Voronoi diagram of four moving sites $p$, $q$, $r$, and $s$.}
     \label{fig:geodesic_voronoi_diagram}
 \end{figure}

Once the metric is fixed, many different problems can be considered. Two of the most basic problems are the computation of
 {\em shortest path maps} and {\em augmented Voronoi diagrams}. A
 shortest path map (or SPM for short) is a partition of the space into
 maximal connected regions so that points in the same region travel in
 the same way to the fixed
 source~\cite{guibas1987spm,hershberger1999sssp}.

 Similar to shortest paths, there are several ways in which paths can be considered ``in the same way''. For the purposes of this paper, we use the well established {\em combinatorial} approach: recall that geodesics are simple polygonal chains, thus $\geod(s,t)$ is described as $\geod(s,t)=(s=v_0,v_1, \ldots,v_k,t=v_{k+1})$ the union of segments $\overline{v_0v_1}, \overline{v_1v_2}, \ldots, \overline{v_kv_{k+1}}$ for some $k\geq 0$. Two paths $\geod(s,t)=(s=v_0,v_1, \ldots,v_k,t=v_{k+1})$ and $\geod(s',t')=(s'=w_0,w_1, \ldots,w_{k'},t=w_{k'+1})$ are (combinatorially) equivalent if and only if $k=k'$ and $\{v_1,\ldots, v_k\}=\{w_1,\ldots, w_k\}$. In other words, both paths traverse through the same sequence of intermediate polygon vertices (note that both start and end could be different). Note that if two paths are combinatorially equivalent, then they must traverse the same vertices in either the same or reverse order (i.e., the vertex order cannot be rearranged). This holds true because subpaths of shortest paths are always shortest paths. For a full proof of this statement (and more interesting properties of geodesics) we refer the interested reader to the seminal paper by Guibas {\em et al.}~\cite{guibas1987spm}. From now on, for simplicity in the description, we assume that the vertices of $P$ are in general position. That is, any segment contained in the closure $P$ passes through at most two vertices. This assumption can be removed using standard data perturbation techniques.

Voronoi diagrams are another fundamental object of computational geometry. Given a collection of sites inside some metric space, the Voronoi diagram is a partition of the space into regions so that points in the same region have the same closest site. These diagrams are often used in facility location problems, but there are many applications in different fields of science (see ~\cite{bkos-cgaa-08}, chapter 7 for a survey of this problem and its many applications).

Augmented Voronoi diagrams are a generalization of both shortest path maps and Voronoi diagrams. In our setting, they are defined as follows: given a set $S$ of $n$ points (which from now on we call {\em sites}) inside a simple polygon $P$ of $m$ vertices, the {\em augmented Voronoi diagram} of $P$ with respect to $S$ is a subdivision of $P$ into cells so that points $p,q$ in the same cell $c$ have the same closest site and the paths from $p$ and $q$ to the nearest site are combinatorially equivalent (see Fig.~\ref{fig:geodesic_voronoi_diagram} for an illustration).

These diagrams often come naturally in different settings. For example, in the previously mentioned robotics setting where $P$ represents the space that the robot can travel to, the sites could be some service location (say, battery stations). Robots move along $P$ doing some task, but when they need recharging, they must travel to the nearest charging station. The shortest path can be found with the help of the augmented Voronoi diagram. The main advantage of augmented diagrams is that full paths need not be stored: from the each cell we only need to know the first vertex of the shortest path to its closest site. Once we reach that vertex, we are at a different cell of the augmented Voronoi diagram, so we query the diagram and obtain a new intermediate vertex for the robot. This process is repeated until the robot eventually reaches the charging station.

In this paper we consider another natural extension of these fundamental problems: rather than considering static sites, we want to study the case in which sites can move. In the robotics example, two agents might be trying to meet, or one agent might want to evade
the other, or one agent might simply need to meet up with a second one that is performing a
different task~\cite{lubiw2017cops}. The existing static algorithms would need to recompute the
diagram after each infinitesimal movement. Instead, in this paper we aim to maintain as much information as possible, doing only local changes whenever strictly necessary. A data structure that can
handle such a setting is known as a {\em kinetic data structure} (or KDS for
short)~\cite{bgh-dsmd-99}.

\subsection{Related Work}
In this paper we combine three fundamental properties of the computational geometry community: polygons, Voronoi diagrams, and kinetic data structures.
Surprisingly, there is very little work that combines the three results.

Aronov was the first to apply the concept of augmented Voronoi diagram
to geodesic environments~\cite{aronov1989geodesic}. This structure has proven to be of critical importance for obtaining efficient solutions to other related problems such as finding center points, closest pairs, nearest neighbors, and constructing spanners~\cite{oh2018_2center,papadopoulou1998geodesic}.
Aronov proved that the augmented Voronoi diagram $\VD_P(S)$ of a set $S$ containing $n$ static point
sites in a simple polygon $P$ of $m$ vertices has complexity $O(n+m)$. 
Moreover, he presented an $O((n+m)\log (n+m)\log n)$ time algorithm for
constructing $\VD_P(S)$, which was improved to $O((n+m)\log (n+m))$ by
Papadopoulou and Lee~\cite{papadopoulou1998geodesic}. Recently, there have been
several improved algorithms~\cite{liu2018near_optimal_gvd, oh_ahn2017voronoi_journal}
which ultimately lead to an optimal $O(m + n\log n)$ time algorithm by
Oh~\cite{oh2019voronoi}. Furthermore, Agarwal~\etal\cite{dyn_geod_nn2018}
recently showed that finding the site in $S$ closest to an arbitrary
query point $q \in P$ --- a key application of geodesic Voronoi diagrams --- can
be achieved efficiently even if sites may be added to, or removed from, $S$.

There are no known results on maintaining an (augmented) geodesic
Voronoi diagram when multiple sites $S$ move continuously in a simple
polygon $P$. In case there is only one site $s$, Aronov
\etal~\cite{agtz-vqmsp-02} presented a KDS that maintains the shortest
path map $\SPM_s$ of $s$. Their data structure uses $O(m)$ space, and
processes a total of $O(m)$ events in $O(\log m)$ time
each\footnote{The original description by Aronov
  \etal~\cite{agtz-vqmsp-02} uses a dynamic convex hull data structure
  that supports $O(\log^2 m)$ time queries and updates. Instead, we
  can use the data structure by Brodal and Jacob~\cite{brodal02dynam} which
  supports these operations in $O(\log m)$ time.}. Karavelas and
Guibas~\cite{karavelas2001kinetic_cdt}, provide a KDS to maintain a
constrained Delaunay triangulation of $S$. This allows them to
maintain the geodesic hull of $S$ with respect to $P$, and the set of nearest
neighbors in $S$ (even in case $P$ has holes). Their KDS processes
$O((m+n)^3\beta_ z(n+m))$ events in $O(\log (n+m))$ time each, where
$\beta_z(n)=\lambda_z(n)/n$ and $\lambda_z(n)$ is the maximum length
of a Davenport-Schinzel sequence of $n$ symbols of order
$z$~\cite{sharir1995davenport}. Here, and
throughout the rest of the paper $z$ denotes some small (natural) constant that
depends on the algebraic degree of the polynomials describing the
movement of the sites. Note that, for any constant $z$,
$\lambda_z(n)$ is near linear, and thus $\beta_z(n)$ is near-constant.

Parallel to this work on geodesic environments, kinetic data structures (KDS) have been used for a wide range of problems in different settings. We refer the reader to the survey by Basch \etal~\cite{bgh-dsmd-99} for an overview of these results. Our data structures follows the same framework: points move linearly in a known direction.
Each KDS maintains a set of \emph{certificates} that together certify that the KDS currently correctly represents the target structure. Typically these certificates involve a few objects each and
represent some simple geometric primitive. For example a certificate
may indicate that three points form a clockwise oriented triangle. As
the points move these certificates may become invalid, requiring the
KDS to update. This requires repairing the target structure and
creating new certificates. Such a certificate failure is called an
(\emph{internal}) \emph{event}. An event is \emph{external} if the
target structure also changes. The performance of a KDS is measured
according to four measures. A KDS is considered \emph{compact} if it
requires little space, generally close to linear, \emph{responsive} if
each event is processed quickly, generally polylogarithmic time,
\emph{local} if each site participates in few events, and
\emph{efficient} if the ratio between external and internal events is
small, generally polylogarithmic. Note that for efficiency it is
common to compare the worst-case number of events for either case.

Guibas \etal~\cite{guibas1991kineticvd} studied maintaining
the Voronoi diagram in case $P=\R^2$ and distance is measured by the
Euclidean distance. They prove that the combinatorial structure of $\VD_{\R^2}(S)$ may change
$\Omega(n^2)$ times, and present an a KDS that handles at most
$O(n^3\beta_4(n))$ events, each in $O(\log n)$ time. Their results actually extend to
slightly more general types of movement. It is one of the long
outstanding open problems if this bound can be
improved~\cite{topp,handbook04}. Only recently,
Rubin~\cite{rubin2015kdt} showed that if all sites move linearly and
with the same speed, the number of changes is at most $O(n^{2+\eps})$
for some arbitrarily small $\eps > 0$. For arbitrary speeds, the best
known bound is still $O(n^3\beta_4(n))$. When the distance function is
specified by a convex $k$-gon the number of changes is
$O(k^4n^2\beta_z(n))$~\cite{agarwal2015kvd_convex}.

\subsection{Results and paper organization}

We present the first KDS to maintain the augmented geodesic Voronoi diagram $\VD_P(S)$ of a set
$S$ of $n$ sites moving linearly inside a simple polygon $P$ with $m$ vertices. Our
results provide an important tool for maintaining related structures in which
the agents (sites) move linearly within the simple polygon (e.g. minimum
spanning trees, nearest-neighbors, closest pairs, etc.).

To this end, we prove a tight $O(m^3)$ bound on the number
of combinatorial changes in a single bisector, and develop a compact,
efficient, and responsive KDS to maintain it
(Section~\ref{sec:Bisector}). Our KDS for the bisector uses $O(m)$
space and processes events in $O(\log^2 m)$ time. We then show that the
movement of the Voronoi center $c_{pqs}$ ---the point equidistant to
three sites $p,q,s \in S$--- can also change $O(m^3)$ times
(Section~\ref{sec:center}). We again show that this bound is tight,
and develop a compact, efficient, and responsive KDS to maintain
$c_{pqs}$. The space usage is linear, and handling an event takes
$O(\log^2 m)$ time. Both our KDSs can be made local as well, and
therefore efficiently support updates to the movement of the
sites. Building on these results we then analyze the full Voronoi
diagram $\VD_P(S)$ of $n$ moving sites
(Section~\ref{sec:Voronoi_diagram}). We identify the different types
of events at which $\VD_P(S)$ changes, and bound their
number. Table~\ref{tab:voronoi_events} gives an overview of our
bounds. We then develop a compact KDS to maintain
$\VD_P(S)$.

\begin{table}[tbh]
  \centering
  \begin{tabular}{p{.3\textwidth}p{.25\textwidth}p{.3\textwidth}}
    \toprule
    Event                         & Lower bound            & Upper bound \\
    \midrule
    $1,2$-collapse/expand         & $\Omega(m^2n)$          & $O(m^2n^2)$ \\
    $1,3$-collapse/expand         & $\Omega(mn\min\{n,m\})$ & $O(m^2n^2 \min\{m\beta_z(n),n\})$ \\
    $2,2$-collapse/expand         & $\Omega(m^3n)$          & $O(m^3n\beta_6(n))$\\
    $2,3$-collapse/expand         & $\Omega(mn^2+m^3n)$     & $O(m^3n^2\beta_6(n)\beta_z(n))$\\
    $3,3$-collapse/expand         & $\Omega(mn^2 + m^2n)$   & $O(m^3n^3\beta_z(n))$ \\
    vertex                        & $\Omega(m^2n)$          & $O(m^2n\beta_6(n))$\\
    \bottomrule
  \end{tabular}
  \caption{The different types of events at which the geodesic Voronoi diagram
    changes, and their number. At an $a,b$-collapse event two vertices of
    $\VD_P(S)$ with degrees $a$ and $b$ collide and one disappears. Similarly,
    at an $a,b$-expand one such a vertex appears. At a vertex event a
    vertex of $\VD_P(S)$ collides with a vertex of $P$.}
  \label{tab:voronoi_events}
\end{table}

\section{Preliminaries}
\label{sec:prelim}
Let $P$ be a simple polygon with $m$ vertices, and let $s,q\in P$. Let $\geod(s,q)$ be
the shortest path between $s$ and $q$ that stays entirely inside $P$ (we view $P$ as a closed polygon and thus such path is known to always exist and to be unique~\cite{m-spaop-93}).
We measure length of a path by the sum of the Euclidean edge
lengths. Such a shortest path is known as a geodesic, and its length
as the \emph{geodesic distance} and is denoted by $\geodlen{s,q}$. From now on, for simplicity in the expression we remove the term {\em geodesic} when talking about distances. Thus, any term that depends on an implicit distance (such as {\em closer} or {\em circle}) refers to the geodesic counterparts (i.e., geodesically closer or geodesic circle).

Let $\Time \subseteq \R$ denote the time domain, and let $S$ be a set
of $n$ point sites that each move with a fixed speed and direction in
a space $P$. That is, we see each point $s \in S$ as a linear function
from $\Time$ to $P$. More precisely, let
$I_s = \{ t \mid s(t) \text{ lies inside } P\} \subseteq \R$ expresses
the time interval during which entity $s$ moves inside $P$ then
$\Time = \cap_{s \in S} I_s$. In the remainder of the paper, we will
not distinguish between a function and its graph.  The \emph{Voronoi
  diagram} $\VD_P(S)$ of $S$ is a partition of $P$ into $n$ regions,
one per site $s \in S$, such that any point $q$ in such a region
$\V_s$ is closer to $s$ than to any other site in $S$. Note that the
Voronoi diagram also changes with time (and thus technically
$\VD_P(S)=\VD_P(S,t)$), but we omit the dependency of $t$ for
simplicity in notation.

\paragraph{General Position}\label{sec:generalpos}

In static domains (i.e., when points do not move), it is common to assume that the input satisfies some form of {\em general position} (say, there are no three collinear input vertices). These general position assumptions can often be removed using standard symbolic perturbation techniques (imagine performing an infinitesimally small random perturbation to the input values; we refer the interested reader to~\cite{degeneracy} for more details on how to implement this without having to modify the input).

Let $V=V(t)$ to denote the union of the set of sites $S=S(t)$ and vertices
$P$ (recall that vertices are static, thus there is no dependency on
$t$).
Along the paper we would like to make the following assumptions:
\begin{enumerate}
\item No line contains more than two points of $V$. 
\item For any $p \in P$ there do not exist more than three of $V$ that
  are geodesically equidistant to $p$. 
\item For any $p\in V$ there do not exist two distinct points of $V$ that are geodesically equidistant to $p$.

\end{enumerate}

Each of these constraints can be expressed with algebraic equations of
constant degree, thus they can be achieved on a static set of points
using symbolic perturbation. Unfortunately, the same cannot be said
when sites move along time: for example, when a point moves fast
through the area between two slowly moving points, at some point in
time the moving point will align with the other two. In this case,
symbolic perturbation can help to split and limit the duration of
these degeneracies.

To make this more precise we define the concept of
a \emph{singular exception}. Given a set of moving sites $S$ in a
simple polygon $P$, we say that a series of algebraic constraints on
$V=V(t)$ are satisfied with singular exceptions if $(a)$ all
constraints are satisfied for all $t\in \R$ except a finite number of
exceptions $t_1, \ldots t_k$, $(b)$ for each $t_i$ there is exactly
one constraint that is not satisfied and there is exactly one instance
that does not satisfy the constraint (say, at $t_1$ we have exactly
one line containing more than two points of $V$) and, $(c)$ the
constraint being violated has exactly one additional point above the
allowed limit (i.e., the line of $t_1$ passes through exactly three
points of $V$).

By viewing $V=V(t)$ as lines in three dimensions, the assumptions above with singular exceptions can be expressed algebraically, and thus obtained using symbolic perturbation. As mentioned above, an infinitesimally small but random permutation on the input would guarantee this with high probability (but there is no need to actually modify the input~\cite{degeneracy}). We note that along the paper we will introduce additional general position assumptions. For simplicity in the description all of them will be with singular exceptions.


\paragraph{Geodesic Voronoi Diagrams}

\begin{figure}[tb]
    \centering
    \includegraphics{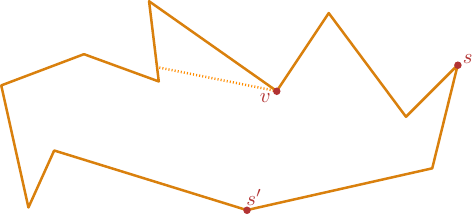}
    \caption{The extension segment $E_{vs}$ (shown as a dashed segment). Note that $E_{vs'}$ is the degenerate segment from $v$ to $v$.}
    \label{fig:extension}
  \end{figure}

We now review some known properties of geodesic Voronoi diagrams and
shortest path maps that we will use. Let $\SPM_s$ be the shortest path
map of $s$. For all points in a single region of $\SPM_s$, the shortest
path from $s$ has the same internal vertices. Each such region $R$ is
star-shaped with respect to the last internal vertex $v$ on the
shortest path. Often it will be useful to refine $R$ into triangles
incident to $v$ (by adding segments between $v$ and other vertices of $P$ in the same region). We refer to the resulting subdivision of $P$ as the
\emph{extended} shortest path map\footnote{In order to have a proper subdivision, we would need to consider one or even zero dimensional regions in which points have two or three shortest paths to $s$. For simplicity, we follow a slight abuse done in previous papers and consider each region as closed with nonempty intersection with neighboring cells. In this way, a point with two or three shortest paths to $s$ belongs to two or three regions at the same time. This is not a proper subdivision, but simplifies things from a computational perspective.}. With some abuse of notation we will
use $\SPM_s$ to denote this subdivision as well. Fixed a source $s$ and a reflex vertex $v$, the \emph{extension segment}
$E_{vs}=E_v$ is defined as follows: shoot a ray from $v$ that is parallel to the last edge in $\geod(s,v)$ and goes away from $v$. Extend that ray until it properly intersects with $P$. Note that this segment could degenerate to a point (see Figure~\ref{fig:extension}). When $E_v$ is not degenerate it
splits $P$ into two regions, one of which contains $s$. More interestingly, for all points in the other region their shortest path to $s$ passes through $v$.

Given two sites $p$ and $q$, the bisector $B_{pq}$ is the set of all
points that are equidistant to $p$ and $q$. If no vertex of $P$ lies
on the bisector, then $B_{pq}$ is a continuous curve connecting two
points on $\partial P$. Moreover, $B_{pq}$ can be decomposed into $O(m)$ pieces, each of which is a subarc of a
hyperbola (that could degenerate to a line segment
\cite{aronov1989geodesic,Mitchell1991}). More generally, for sites in
general position we have that:

\begin{lemma}[Aronov~\cite{aronov1989geodesic}]
  \label{lem:structure_gvd}
  $\VD_P(S)$ consists of $O(n)$ vertices with
  degree 1 or 3, and $O(m)$ vertices of degree 2. For each degree 2
  vertex $v$ there is are $p,q \in S$ so that $v$ lies on the bisector $B_{pq}$
  and $v$ lies on extension segment of $\SPM_p$ or $\SPM_q$. All edges of $\VD_P(S)$
  are hyperbolic arc segments. Every vertex $v$ of $P$ contributes at most one
  extension segment $E_v$ to $\VD_P(S)$.
\end{lemma}

\begin{lemma}[Aronov \etal~\cite{agtz-vqmsp-02}]
  \label{lem:kinetic_spm}
  Let $s$ be a point moving linearly inside a simple polygon $P$ with $m$
  vertices. The extended shortest path map $\SPM_s$ changes at most $O(m)$
  times.
\end{lemma}

\begin{lemma}
  \label{lem:closest_to_vertex}
  Let $v$ be a vertex of $P$, there are $O(mn\beta_6(n))$ time intervals in
  which $v$ has a unique closest site $s \in S$, and the distance from $v$ to
  $s$ over time is a hyperbolic function.
\end{lemma}

\begin{proof}
  For every site $s \in S$, consider the distance function
  $f_s(t)=\geodlen{s(t),v}$ from $s$ to $v$. The site closest to $v$
  corresponds to the lower envelope of these $n$ functions. Each
  function $f_s(t)$ consists of $O(m)$ pieces, each of which is of the
  form $\|vu\|+\|us(t)\|$, for some vertex $u$ of $P$. Since $s(t)$ is
  a linear function in $t$, we will argue that two such pieces can
  intersect at most four times. It then follows that the lower envelope has
  complexity $O(mn\beta_6(n))$~\cite{sharir1995davenport}.

  Let $s$ and $r$ be entities moving linearly, each with a constant
  speed, and consider a time interval $I$ during which $u$ is the last
  vertex on $\geod(v,s(t))$, and $w$ is the last vertex on
  $\geod(v,r(t))$. We have to argue that within time interval $I$, the
  functions $\geodlen{v,s(t)}$ and $\geodlen{v,r(t)}$ intersect at
  most four times, hence that $\geodlen{v,s(t)} - \geodlen{v,r(t)}$
  has at most four roots. We use that $s(t)$ and $r(t)$ are linear
  functions in $t$, and that in this interval $I$, we have that
  $\geodlen{v,s(t)} = \|s(t)u\| + \geodlen{u,v} = \sqrt{(s(t)_x-u_x)^2
    + (s(t)_y-u_y)^2} + \geodlen{u,v}$ and
  $\geodlen{v,r(t)} = \|r(t)w\| + \geodlen{v,w} = \sqrt{(r(t)_x-w_x)^2 +
  (r(t)_y-w_y)^2} + \geodlen{w,v}$. By repeated squaring we now obtain
  a polynomial of degree four. Hence, there are at most four
  roots. The lemma follows.
\end{proof}

\section{A Single Bisector}
\label{sec:Bisector}

\begin{figure}[tbh]
  \centering
  \includegraphics{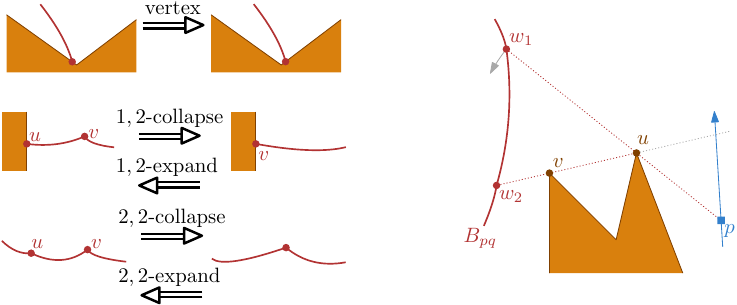}
  \caption{ (left) The types of events at which the structure of
    $B_{pq}$ changes. (right) An example of a $2,2$-collapse event. As
    $p$ moves, bisector vertex $w_1$ rotates around polygon vertex
    $u$. When $p$ becomes collinear with vertices $u$ and $v$, it creates a
    $2,2$-collapse where $w_1$ collides with neighboring bisector
    vertex $w_2$.}
  \label{fig:bisector_events}
\end{figure}

In this section we study the single bisector of a fixed pair of sites $p$ and $q$ and study how it changes as the points move linearly. Let $b_{pq}(t)$ and $b_{qp}(t)$
be the endpoints of the bisector $B_{pq}$ defined so that $p$ lies to
the right of $B_{pq}(t)$ when following the bisector from $b_{pq}(t)$
to $b_{qp}(t)$. As $p$ and $q$ move the structure of $B_{pq}$ changes
at discrete times, or events. We distinguish between the following
types of events (see Fig.~\ref{fig:bisector_events}, left):
\begin{itemize}[nosep]
\item \emph{vertex} events, at which an endpoint of $B_{pq}$ coincides with a
  vertex of $P$,
\item \emph{$1,2$-collapse} events, at
which a degree 2 vertex (an interior vertex) of $B_{pq}$ disappears as it
collides with a degree 1 vertex (an endpoint),
\item \emph{$1,2$-expand} events, at which a new degree 2 vertex appears from
  a degree 1 vertex,
\item \emph{$2,2$-collapse} events at which a degree 2 vertex disappears by
  colliding with an other degree 2 vertex, and
\item \emph{$2,2$-expand} events, at which a new degree 2 vertex appears from
  a degree 2 vertex.
\end{itemize}

\begin{figure}[tb]
  \centering
  \includegraphics{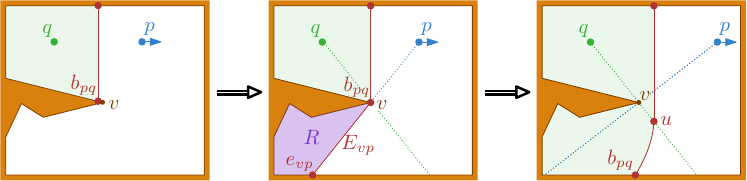}
  \caption{A vertex event at $v$ may coincide with a $1,2$-expand
    event. At the time of the event all points in $R$ are equidistant
    to $p$ and $q$, and $b_{pq}$ jumps from $v$ to $e_{vp}$.}
  \label{fig:kds_bisector_split}
\end{figure}

We now briefly justify that these are all the events that can happen
to a single bisector. When the sites are in general position,
Aronov~\cite[Lemma 3.2]{aronov1989geodesic} showed that the bisector
is the concatenation of hyperbolic arcs and line segments that touches
the boundary of $P$ in two points. Thus, the vertices of the bisector
are of degree at most two, and therefore, our characterization of
$d_i,d_j$-collapse and expand events in terms of the degrees $d_i$ and
$d_j$ of the bisector vertices captures all changes in the interior of
the polygon. Furthermore, since the bisector touches the boundary of
the polygon only in its two endpoints, the events where the bisector
changes due to the polygon boundary are when the endpoint of the
bisector coincides with a vertex of the polygon; i.e. at vertex
events. 
Note that, even with general position assumptions, multiple events could happen at the same time and place.
An example of this is shown in Fig.~\ref{fig:kds_bisector_split}: the bisector passes through a reflex corner and thus ``jumps'' from an edge to another.
This is in fact the combination of a vertex event and a $1,2$-expand event. Each time we process an event we check if multiple events are happening, and if so we treat them one at a time.

In our analysis, we will be counting the number of events of each type
separately. Note that therefore we are actually double-counting
simultaneous events like the one in
Fig.~\ref{fig:kds_bisector_split}. In
Section~\ref{sub:Bounding_the_Number_of_Events_bisec} we prove that
there are at most $O(m^2)$ vertex and $1,2$-collapse events, and at
most $O(m^3)$ $2,2$-collapse events. The number of expand events can
be similarly bounded. Despite our double-counting, we can show that
our resulting $O(m^3)$ bound on the number of combinatorial changes of
$B_{pq}$ is tight in the worst case. In
Section~\ref{sub:A_KDS_for_maintaining_the_Bisector} we then argue
that there is a KDS that can maintain $B_{pq}$ efficiently.

\subsection{Bounding the Number of Events}
\label{sub:Bounding_the_Number_of_Events_bisec}

We start by showing that the combinatorial structure of a bisector may
change $\Omega(m^3)$ times. We then argue that there is also an
$O(m^3)$ upper bound on the number of such changes.

\begin{lemma}
  \label{lem:lowerbound_bisector}
  The combinatorial structure of the bisector $B_{pq}(t)$ can change $\Omega(m^3)$ times.
\end{lemma}

\begin{proof}
  The main idea is to construct two chains of reflex vertices, $C_p$
  and $C_q$, each of complexity $\Omega(m)$, and such that their
  extension segments (from $\SPM_p$ and $\SPM_q$, respectively) define
  a grid of complexity $\Omega(m^2)$. We then make the bisector
  $B_{pq}$ sweep across this grid $\Omega(m)$ times using two
  additional chains of reflex vertices $D_p$ and $D_q$. See
  Fig.~\ref{fig:lowerbound_22collapse_wineglass}(a) for an
  illustration. Each time $B_{pq}$ sweeps across an intersection
  point, this causes a combinatorial change, and thus $B_{pq}$ goes
  through $\Omega(m^3)$ combinatorial changes in total.

  The vertices in the chain $C_p$ are almost collinear, so that the
  region containing the grid of extension segments is tiny. Chain
  $C_q$ is a mirrored copy of $C_p$. Let $u$
  and $v$ be two consecutive vertices in $C_p$, let $w$ and $z$ be two
  consecutive vertices in $C_q$, and consider a time $t$ at which $v$
  and $w$ define a segment $B^{vw}$ of the bisector
  $B_{pq}(t)$. Hence, for any point $b$ on $B^{vw}$ we have
  $\geodlen{p(t),v} + \|vb\| = \geodlen{b,p(t)} =
  \geodlen{b,q(t)}$. See
  Fig.~\ref{fig:lowerbound_22collapse_wineglass}(b). Once the bisector
  sweeps over the intersection point of $E_v$ and $E_z$ (from right to
  left; as $q$ is getting closer to the intersection point), this
  segment $B^{vw}$ collapses into a point, and gets replaced by a segment
  defined by $u$ and $z$ (i.e. points equidistant to $p$ and $q$ for
  which the shortest paths go have $u$ and $z$ as last vertex,
  respectively). Hence, the intersection point corresponds to a
  combinatorial change in $B_{pq}$.

 \begin{figure}[tbh]
   \centering
   \includegraphics{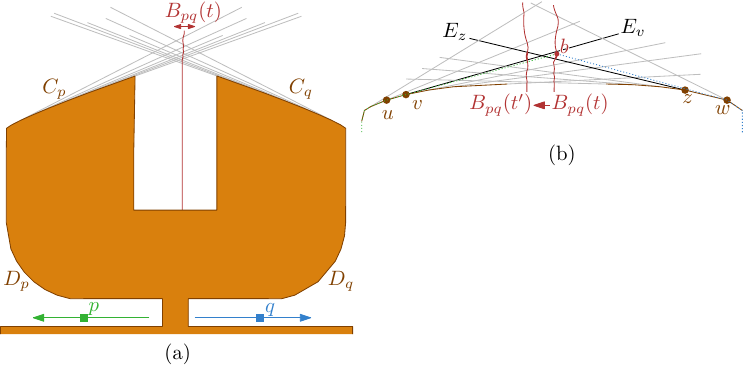}
   \caption{The bisector $B_{pq}$ may be involved in $\Omega(m^3)$
     $2,2$-collapse events. (a) A schematic drawing of the relevant
     part of the polygon causing the $\Omega(m^3)$ combinatorial
     changes. (b) When $B_{pq}$ sweeps over an intersection point of
     the two extension segments the structure of the shortest path
     changes, and thus so does the structure of the bisector.  }
   \label{fig:lowerbound_22collapse_wineglass}
 \end{figure}

 Next, we argue that the chains $D_p$ and $D_q$ cause $B_{pq}$ to move
 back and forth across the grid $\Omega(m)$ times. These chains will
 ensure that $p$ and $q$ alternate being the closest to the top of the
 polygon (in particular to the first vertices $v_p$ and $v_q$ of
 chains $C_p$ and $C_q$, respectively). Thus, when $p$ is closest the
 bisector will move to the right and when $q$ is the closest the
 bisector will move to the left. By making $p$ and $q$ move at the
 same speed and having the segments defining the convex chain $D_q$
 start and end in the middle of where the segments of the convex chain
 on $p$'s side ($D_p$) start and end, we can cause this
 alternation. When a vertex $u_i$ from $D_p$ disappears
 from $\geod(p(t),v_p)$, the length of $\geod(p(t),v_p)$ decreases
 more quickly compared to $\geod(q(t),v_q)$ since $\geod(q(t),v_q)$
 still has to go around the copy of $u_i$ in $D_q$.
 When both of these lower convex chains have complexity
 $\Omega(m)$, the bisector sweeps over the $\Omega(m^2)$ middle cells
 $\Omega(m)$ times and thus the combinatorial structure of the
 bisector changes $\Omega(m^3)$ times.
\end{proof}

\begin{lemma}
  \label{lem:vertex_pq}
  The bisector $B_{pq}(t)$ is involved in at most $O(m^2)$ vertex events.
\end{lemma}

\begin{proof}
  At a vertex event one of the endpoints of $B_{pq}(t)$, say $b_{pq}(t)$
  coincides with a polygon vertex $v$. Hence, at such a time
  $\geodlen{p(t),b_{pq}(t)}=\geodlen{q(t),b_{pq}(t)}$. The distance functions from
  $p$ and $q$ to $v$ are piecewise hyperbolic functions with $O(m)$ pieces. So
  there are $O(m)$ time intervals during which both these distance functions
  are continuous hyperbolic functions. A pair of such functions intersects at most
  a constant number of times. Hence, in each such interval there are at most
  $O(1)$ vertex events involving vertex $v$. The bound follows by summing over
  all time intervals and all vertices.
\end{proof}

Fix a polygon vertex $v$, and consider the extension segment
$E_{vp}(t)$ in $\SPM_p(t)$ incident to $v$. Let $e_{vp}(t)$ be the
other endpoint of $E_{vp}(t)$, and observe that, since $E_{vp}(t)$
rotates around $v$, $e_{vp}(t)$ moves monotonically along the boundary
of $P$. That is, as $p$ moves,
$e_{vp}$ moves only clockwise along $\partial P$ or only
counter-clockwise. Hence, the trajectory of $e_{vp}$ consists of
$O(m)$ edges, on each of which $e_{vp}$ moves along an edge of $P$.

\begin{lemma}
  \label{lem:spm_regions_intersected}
  Let $v$ be a vertex of $P$. As $p$ and $q$ move, $e_{vp}$ crosses
  $O(m)$ cells of $\SPM_q$.
\end{lemma}

\begin{proof}
  Consider the restriction of $\SPM_q(t)$ to $\partial P$ as a
  function of $t$, which can be represented as a subdivision \S of the
  two-dimensional space $\Time \times \partial P$. This
  (planar) subdivision has complexity $O(m)$ (by Lemma~\ref{lem:kinetic_spm}). As $p$ moves, the point $e_{pv}$
  traces a curve in this subdivision. The number of $\SPM_p$ cells
  that $e_{pv}$ crosses is thus the number of intersections of this
  curve with the faces of \S. We now argue that there are at most
  $O(m)$ such intersections.

  Observe that the edges of \S trace the trajectories of vertices of
  $\SPM_q(t)$. We distinguish two types of vertices in $\SPM_q(t)$,
  red vertices and blue vertices. The red vertices are either polygon
  vertices, or endpoints $e_{uq}(t)$ of extension segments for which
  $\geod(q(t),e_{uq}(t))$ contains at least one other polygon
  vertex. All other vertices --these correspond to endpoints
  $e_{wq}(t)$ of extension segments such that
  $\geodlen{q(t),e_{wq}(t)}=\|q(t)e_{wq}(t)\|$-- are blue. This coloring
  of the vertices of $\SPM_q$ also induces a coloring of the edges of
  \S. See Fig.~\ref{fig:boundary_spm}. Since all red vertices have
  fixed locations, the corresponding red edges in \S are horizontal
  line segments. Furthermore, observe that every polygon edge is
  visible from $q(t)$ in a single time interval. Hence, there are at
  most two moving endpoints $e_{wq}$ per polygon edge. This implies
  that every horizontal strip defined by two consecutive red edges
  contains at most two blue edges.

  \begin{figure}[tbh]
    \centering
    \includegraphics[clip,trim=0 7cm 0 0]{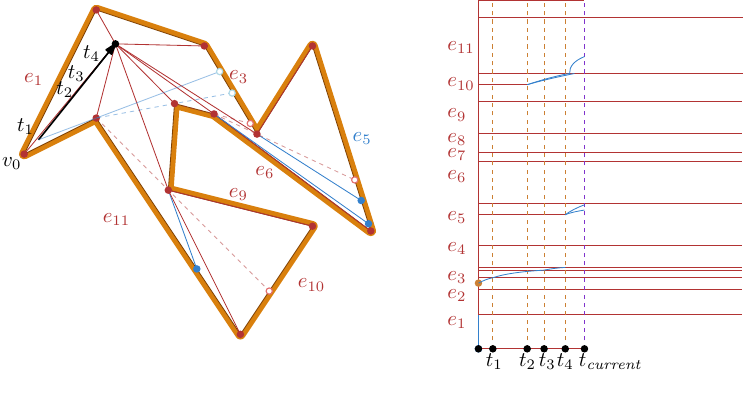}
    \caption{Tracing $\SPM_q(t) \cap \partial P$ as a function of $t$ yields a
      subdivision \S.}
    \label{fig:boundary_spm}
  \end{figure}

  As $p$ moves, the extension segment $E_{pv}$ rotates around $v$, and
  thus the point $e_{pv}$ moves monotonically along the boundary of
  $P$. That is, $e_{pv}$ traces a $t,\lambda$-monotone curve through
  \S (i.e. any line parallel to the $t$-axis or $\lambda$-axis
  intersects it in at most a single connected region). It follows that
  the total number of intersections with the red edges is $O(m)$. We
  further split each horizontal strip (defined by the red edges of \S)
  at vertices of the curve traced by $e_{pv}$. Since this curve has
  complexity $O(m)$, the total number of strips remains $O(m)$. Each
  such strip is (still) crossed by at most two blue edges. The edges
  in the trajectory of $e_{pv}$ as well as those blue edges are curves
  of constant algebraic degree. Hence a pair of such curves intersect only
  $O(1)$ times, and thus every strip contributes at most a constant
  number of intersections. Since we have $O(m)$ strips, we thus also
  get at most $O(m)$ intersections. The lemma follows.
\end{proof}

\begin{lemma}
 \label{lem:12_collapse_pq}
 The bisector $B_{pq}(t)$ is involved in at most $O(m^2)$ $1,2$-collapse events.
\end{lemma}

\begin{proof}
 Fix a vertex $v$, and consider the endpoint $e_{vp}(t)$ of $E_{vp}(t)$. By
 Lemma~\ref{lem:spm_regions_intersected} this point intersects at most $O(m)$
 regions of $\SPM_p$ and $\SPM_q$ throughout the motion of $p$ and $q$. It
 then follows that there are $O(m)$ time intervals during which the distances
 $f(t) = \geodlen{p(t),e_{vp}(t)}$ and $g(t)=\geodlen{q(t),e_{vp}(t)}$ are both
 continuous and constant algebraic degree. We restrict the domains of $f$ and $g$ to
 the time intervals during which $p$ is closer to $v$ than $q$. It follows
 that $f$ and $g$ still consist of $O(m)$ pieces.

 Now observe that any $1,2$-collapse event of $B_{pq}(t)$ on $E_{vp}(t)$
 corresponds to a time where: $v$ is closer to $p$ than $q$, and the endpoint
 $e_{vp}(t)$ of $E_{vp}(t)$ is equidistant to $p$ and $q$, that is,
 $f(t) = g(t)$. Hence, the number of such events equals the number of
 intersections between (the graphs of) $f$ and $g$. Since $f$ and $g$ both
 consist of $O(m)$ pieces, each of constant algebraic degree, the number of
 intersections, and thus the number of $1,2$-collapse events on $E_{vp}(t)$ is
 $O(m)$. Similarly, the number of events on $E_{vq}(t)$  is $O(m)$. The lemma
 follows by summing these events over all vertices $v$.
\end{proof}

\begin{lemma}
 \label{lem:22collapse_upperbound_pq}
 The bisector $B_{pq}(t)$ is involved in at most $O(m^3)$ $2,2$-collapse events.
\end{lemma}

\begin{proof}
 We observe that in a $2,2$-collapse of edge $(a,b)$ of $B_{pq}(t)$ both $a$
 and $b$ must be on extension segments $E_u$ and $E_v$ in the shortest path
 map of $p$ or $q$ at time $t$. Hence a $2,2$-collapse occurs at the
 intersection point of $E_u$ and $E_v$. In particular, at such an event, the
 distances from $p$ to $a$ and from $q$ to $a$ are equal.

 If $E_u$ and $E_v$ both occur in a single shortest path map, say $\SPM_p(t)$,
 this $2,2$-collapse event corresponds to an event at which the combinatorial
 structure of $\SPM_p$ changes. Thus, the total number of such changes is at
 most $O(m)$~(Lemma~\ref{lem:kinetic_spm}). We thus focus on the case that
 $E_u$ is an extension segment in $\SPM_p$ and $E_v$ is an extension segment
 in $\SPM_q$.

 Since the combinatorial structure of $\SPM_p$ and $\SPM_q$ changes at most
 $O(m)$ times, the total number of pairs of extension segments that we have to
 consider is $O(m^2)$. We now argue that for each such pair there are at
 most $O(m)$ times $t$ where $\geodlen{p(t),a(t)}=\geodlen{q(t),a(t)}$, and thus
 there are at most $O(m)$ $2,2$-collapse events involving the pair
 $(E_u,E_v)$. The total number of $2,2$-collapse events is then $O(m^3)$ as
 claimed.

 The distance function from $p$ to $u$ is a piecewise hyperbolic function with
 $O(m)$ pieces. The same is true for the distance function from $q$ to $v$. We
 then consider maximal time intervals during which both these distance
 functions are continuous, and during which $E_u$ and $E_v$ are part of their
 respective shortest path maps. There are at most $O(m)$ such intervals. Since
 $E_u$ and $E_v$ move along a trajectory of constant complexity (i.e. they
 either rotate continuously around $u$ and $v$, respectively, or remain
 static), the distance function $f(t,\mu) = \geodlen{p(t),u} + \|uE_v(\mu)\|$
 from $p$ via $u$ to a point $E_v(\mu)$ on $E_v$ also consists of $O(m)$
 pieces, each of constant algebraic degree. The same applies for the distance
 function $g(t,\lambda) = \geodlen{q(t),v}+\|vE_u(\lambda)\|$, for $E_u(\lambda)$
 on $E_u$. Therefore, during each time interval, the distance functions from
 $p$ to $a$ and from $q$ to $a$ are also continuous low-degree algebraic
 functions. Such functions intersect at most $O(1)$ times, and thus the number
 of $2,2$-collapse events in every interval is at most constant. Since we have
 $O(m)$ intervals, the number of $2,2$-collapses involving $E_u$ and $E_v$ is
 $O(m)$.
\end{proof}

\begin{theorem}
  \label{thm:upperbound_bisector}
  Let $p$ and $q$ be two points moving linearly inside $P$. The
  combinatorial structure of the bisector
  $B_{pq}$ of $p$ and $q$ can change $O(m^3)$ times. This bound is tight in the
  worst case.
\end{theorem}

\begin{proof}
  The combinatorial structure of the bisector $B_{pq}$ changes either at a vertex event, a $1,2$-collapse, or
  a $2,2$-collapse. By Lemmas~\ref{lem:vertex_pq},~\ref{lem:12_collapse_pq},
  and~\ref{lem:22collapse_upperbound_pq} there are at most most $O(m^2)$,
  $O(m^2)$, and $O(m^3)$, such events respectively. We can use symmetric
  arguments to bound the expand events. Thus the structure of $B_{pq}$ changes at most $O(m^3)$
  times. By Lemma~\ref{lem:lowerbound_bisector} this bound is tight in the
  worst case.
\end{proof}

Even though the (combinatorial structure of the) entire bisector
$B_{pq}$ may change $O(m^3)$ times in total, the trajectories of its
intersection points with the boundary of $P$ have complexity at most
$O(m^2)$:

\begin{lemma}
  \label{lem:complexity_bpq}
  The trajectory of $b_{pq}$ has $O(m^2)$ edges, each of which is a
  constant degree algebraic curve.
\end{lemma}

\begin{proof}
  The trajectory of $b_{pq}$ changes only at vertex events or at $1,2$-collapse
  events. By Lemmas~\ref{lem:vertex_pq} and~\ref{lem:12_collapse_pq} the number
  of such events is at most $O(m^2)$. Fix a time interval in between two
  consecutive events, and assume without loss of generality that $b_{pq}(t)$
  moves on an edge of $P$ that coincides with the $x$-axis. We thus have
  $b_{pq}(t)=(x(t),0)$, for some function $x$. Since
  $\geodlen{p(t),b_{pq}(t)}=\geodlen{q(t),b_{pq}(t)}$ we have that
  $\sqrt{Q_p(t)} + C_p + \sqrt{(x(t)-D_p)^2+E_p} = \sqrt{Q_q(t)} + C_q +
  \sqrt{(x(t)-D_q)^2+E_q}$, for some quadratic functions $Q_p(t)$ and $Q_q(t)$
  and constants $C_p,C_q,D_p,D_q,E_p$, and $E_q$. By repeated squaring and
  basic algebraic manipulations it follows that $x(t)$ is some constant degree
  algebraic function in $t$. Hence, every edge in the trajectory of $b_{pq}$
  corresponds to a constant degree algebraic curve.
\end{proof}

\subsection{A Kinetic Data Structure to Maintain a Bisector}
\label{sub:A_KDS_for_maintaining_the_Bisector}

We first describe a simple, yet naive, KDS to maintain $B_{pq}$ that
is not responsive and then show how to improve it to obtain a
responsive KDS.

\subsubsection{A Non-Responsive KDS to Maintain a Bisector}
\label{ssub:Naive_KDS_Bisector}

Our naive KDS for maintaining $B_{pq}$ stores: (i) the extended shortest
path maps of $p$ and $q$ using the data structure of
Aronov~\etal~\cite{agtz-vqmsp-02}, (ii) the vertices of $B_{pq}$,
ordered along $B_{pq}$ from $b_{pq}$ to $b_{qp}$ in a balanced binary
search tree, and (iii) for every vertex $u$ of $B_{pq}$, the cell of
$\SPM_p$ and of $\SPM_q$ that contains $u$. Since all cells in
$\SPM_p$ and $\SPM_q$ are triangles, this requires only $O(1)$
certificates per vertex. We store these certificates in a priority
queue $\mathcal{Q}$.

At any time where $B_{pq}$ changes combinatorially (i.e. at an event)
the shortest path to a vertex $v$ of $B_{pq}$ changes
combinatorially, which indicates a change in the $\SPM$ cells that contain $v$. Hence, we detect all events.
Conversely, when any vertex $v$ of $B_{pq}$ moves to a different
$\SPM$ cell there is a combinatorial change in the bisector, so each
event triggered by parts (ii) and (iii) of the KDS is an external event.
The events at which $\SPM_p$ or $\SPM_q$ changes are
internal (unless they also cause a combinatorial change in a shortest
path to a vertex of $B_{pq}$).

The events at which $\SPM_p$ or $\SPM_q$ changes are handled as in
Aronov~\etal~\cite{agtz-vqmsp-02}. However, such an event may cause
the shortest path to several bisector vertices to change and we would
need to recompute the certificates for maintaining which cell of the
$\SPM$ each bisector vertex lies in. The internal update for the
$\SPM$ takes $O(\log m)$ time~\cite{agtz-vqmsp-02} and each
certificate can be recomputed in $O(\log m)$ time by computing the
appropriate distance functions. Unfortunately, there may be
$\Theta(m)$ certificates to update, which means such an event may take
$\Theta(m \log m)$ time. We will describe how to avoid this
problem later, but we first describe how the rest of the events of the
KDS are handled.

At any external event, a vertex $u$ of $B_{pq}$ leaves its cell in
$\SPM_p$ or $\SPM_q$, and enters a new one. In all cases we delete the
$O(1)$ certificates corresponding to $u$, and replace them by $O(1)$
new ones. Depending on the type of event, we also update $B_{pq}$
appropriately, i.e. in case of a $1,2$- or $2,2$-collapse event we
remove a vertex from $B_{pq}$ and in case of $1,2$-expand events we
insert a new vertex in $B_{pq}$. We describe how to handle a vertex
event in more detail, as they may happen simultaneously with
$1,2$-collapse or expand events.

Consider a vertex event at vertex $v$ at time $t$, at which a bisector
endpoint, say $b_{pq}$, stops to intersect an edge $\overline{wv}$ of
$\partial P$.

If there are no points in $P$ other than $v$ for which the shortest
path to $p$ or $q$ passes through $v$ then the vertex event is easy to
handle; at such an event $b_{pq}$ simply moves onto the other edge
incident to $v$. In doing so, it crosses into a different cell of
$\SPM_p$ or $\SPM_q$. So, we update the certificates associated with
$b_{pq}$ and continue to the next event.

If there are points $r$ in some region $R \subset P$ for which
$\geod(r,p)$ and $\geod(r,q)$ both pass through $v$, then these points
are now all equidistant to $p$ and $q$, and hence at time $t$ the
entire region $R$ is actually a subset of the bisector $B_{pq}$. See
Fig.~\ref{fig:kds_bisector_split}. This region $R$ is bounded by the
extension segment incident to $v$ in $\SPM_p$, or the extension
segment incident to $v$ in $\SPM_q$, that is, $E_{vp}$ or $E_{vq}$. As
a result, the endpoint $b_{pq}$ will jump to either $e_{vp}$ or
$e_{vq}$ (the other endpoint of $E_{vp}$ or $E_{vq}$,
respectively). Moreover, this extension segment becomes part of the
bisector $B_{pq}$ in the simultaneously occurring $1,2$-expand
event. This new vertex $u$ of $B_{pq}$ moves on the other extension
segment incident to $v$. Hence, to update our KDS we insert a new
vertex $u$ in the balanced binary search tree representing $B_{pq}$,
create the corresponding certificates tracking $u$ in $\SPM_p$ and
$\SPM_q$, and we update the certificates tracking $b_{pq}$ in $\SPM_p$
and $\SPM_q$.

If there are points for which only one of the shortest paths to $p$ or
$q$, say $p$, passes through $v$, the bisector endpoint $b_{pq}$
continues on the other edge incident to $v$, while a new vertex $u$ is
created on $B_{pq}$ moving along $E_{vp}$. We insert $u$ in $B_{pq}$
and create appropriate certificates tracking $u$ and $b_{pq}$ in
$\SPM_p$ and $\SPM_q$ like in the previous case.

Observe that we may also have vertex events at which a bisector
endpoint $b_{pq}$ jumps onto $v$ while it was moving on an edge not
incident to $v$ before in a situation symmetric to in the second case
described above. In such a case the vertex event coincides with a
$1,2$-collapse event in which a bisector vertex $u$ hits $\partial P$
(and thus the boundary of its cell in $\SPM_p$ and $\SPM_q$) at
$v$. This is the reverse situation of the one depicted in
Fig.~\ref{fig:kds_bisector_split}. In this case we delete $u$ and its
certificates, and update the certificates tracking $b_{pq}$.

Each external event involves only a constant number of vertices of
$B_{pq}$. Furthermore, as each such vertex is involved in only a
constant number of certificates. Updating a certificate can easily be
done in $O(\log m)$ time, as this involves a constant number of
updates into the binary search tree representing $B_{pq}$ and the
event queue. Hence, handling an external event can be done in
$O(\log m)$ time.

Observe that at any moment we maintain only $O(m)$ certificates,
stored in a priority queue. We thus use $O(m)$ space, and the
updates to the priority queue require $O(\log m)$ time. The total
number of events for maintaining $\SPM_p$ and $\SPM_q$ is only $O(m)$,
which is dominated by the $O(m^3)$ events at which $B_{pq}$ itself
changes (Theorem~\ref{thm:upperbound_bisector}). So our KDS is
compact, and efficient, but not responsive as updates to the $\SPM$
may require $O(m \log m)$ time. In the next section we show that we do
not actually need to maintain these certificates explicitly.

\subsubsection{A Responsive KDS to Maintain a Bisector}
\label{ssub:bulk_bisector_kds}

\begin{figure}[tbh]
\centering
\includegraphics{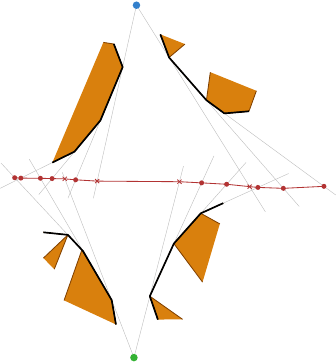}
\caption{A bisector can be split into at most five pieces, here
  separated by degree 2 vertices marked as crosses.
}
\label{fig:bisector_pieces}
 \end{figure}

First we dissect in some more detail the anatomy of a bisector. Each
bisector consists of two endpoints which are degree 1 vertices and a
chain of degree 2 vertices connecting them. We can further divide this
chain based on which parts are directly visible from the sites
defining the bisector. This division results in at most 5 pieces, as
illustrated in Fig.~\ref{fig:bisector_pieces}; some pieces may not be
present in every bisector. First there is a \emph{double-visible}
piece that is visible from both sites $p$ and $q$. Since $P$ is a
simple polygon, this piece consists of a single
line segment. Adjacent to the double-visible piece on either side
there may be a \emph{single-visible} piece that is only visible to $p$
or to $q$, but not both. Lastly, there are up to two
\emph{non-visible} pieces that are not directly visible from either
$p$ or $q$.

 We will still store the bisector vertices in a balanced binary tree
 ordered along the bisector, but we will store the certificates for
 the vertices in each piece separately.

 \subparagraph{Storing the certificates for the partially visible
   pieces} For each of the at most four degree 2 vertices that
 separate the pieces as well as the degree 1 endpoints, we store the
 cells of $\SPM_p$ and $\SPM_q$ that contain it, and track their
 events explicitly as before.

 We observe that for internal vertices of the single-visible piece
 there can be no events. Each of these internal vertices lies on an
 extension segment of a single convex chain of vertices in the simple
 polygon and these extension segments do not intersect. Therefore no
 2,2-collapses can occur, and hence no certificates need to be stored.

 \subparagraph{Storing the ceritificates of the non-visible pieces}
 The non-visible pieces are trickier, since 2,2-collapses may occur
 when a vertex moving on an extension segment of $\SPM_q$ moves to a
 different cell of $\SPM_p$. Fortunately such potential events on a
 single non-visible bisector piece are related and form a strict
 ordering, regardless of the exact distance functions of the various
 vertices to $p$ and $q$.

We define event points to be the locations at which 2,2-collapses that
may occur. Consider two degree 2 vertices $v$ and $w$ that are
internal to a non-visible piece of bisector between sites $p$ and $q$,
such that $v$ and $w$ are adjacent on the bisector and we have that
$v$ is on an extension segment of $\SPM_p$ and $w$ is on an extension
segment of $\SPM_q$. Let the \emph{event point} $\ep_{v,w}$ denote the
intersection between these two extension segments. A 2,2-event between
$v$ and $w$ corresponds to the event point being on the bisector
between $p$ and $q$. Without loss of generality assume that the event
point currently lies in the Voronoi cell of $p$. We can then use the
certificate $\geodlen{\ep_{v,w},p} < \geodlen{\ep_{v,w},q}$ to detect
the 2,2-event between $v$ and $w$. As we saw above maintaining these
certificates explicitly is not efficient as any change in the shortest
path towards $p$ or $q$ requires us to recompute the failure
time. Therefore, we will maintain two balanced binary search trees:
one with all event points that lie in the Voronoi cell of $p$ (ordered
along the bisector), and one with all event points that lie in the
Voronoi cell of $q$. We will augment the balanced binary search trees
so that each node stores some additional information. In particular, a
node $\nu$ in the tree storing the events in the Voronoi cell of $p$
stores: an event point $\ep_\nu$ (in order along the bisector), (ii)
the event point in its sub tree that will be on the bisector first,
and three fields: an \emph{event value}, a \emph{maximum event value},
and two \emph{offsets} (one for each child) that will help maintain
this event point that will be on the bisector first. We will describe
these in more detail later. This way, we end up with a structure
similar to a kinetic tournament~\cite{bgh-dsmd-99}. Therefore, we can
then compute an explicit failure time only for the two event points
stored in the roots of the trees.

\begin{figure}[tbh]
  \centering
  \includegraphics{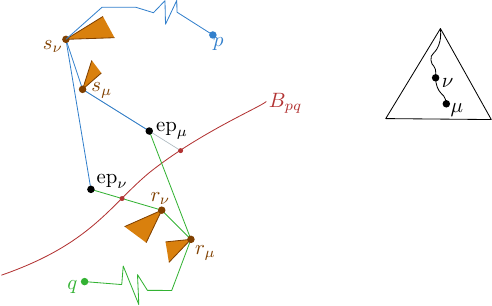}
  \caption{On the left is a schematic drawing of the event points $\ep_\nu$ and $\ep_\mu$
    with their shortest paths towards $p$ and $q$. On the right how
    (the certificates of) $\ep_\nu$ and $\ep_\mu$ are stored in the BST.}
  \label{fig:bisector_kds}
\end{figure}

For a single non-visible bisector piece between sites $p$ and $q$,
Consider event points $\ep_\nu$ and $\ep_\mu$ where $\mu$ is a child
of $\nu$ in the tree. Let $s_\nu$ and $r_\nu$ denote the first polygon
vertex on the shortest path from $\ep_\nu$ towards $p$ and $q$
respectively and let $s_\mu$ and $r_\mu$ be defined symmetrically. See
Fig.~\ref{fig:bisector_kds}. Then we can rewrite the certificate for
$\ep_\nu$ as
\begin{align*}
&\geodlen{\ep_\nu,s_\nu} + \geodlen{s_\nu,p} < \geodlen{\ep_\nu,r_\nu} + \geodlen{r_\nu,q} \\
\equiv~&
\geodlen{\ep_\nu,s_\nu} - \geodlen{\ep_\nu,r_\nu} <  \geodlen{r_\nu,q} -  \geodlen{s_\nu,p},
\end{align*}
and the certificate for $\ep_\mu$ similarly. Then observe that if
$s_\nu = s_\mu$ and $r_\nu = r_\mu$, then $\ep_\nu$ will be on the
bisector before $\ep_\mu$ if and only if
$\geodlen{\ep_\nu,s_\nu} - \geodlen{\ep_\nu,r_\nu} >
\geodlen{\ep_\mu,s_\mu} - \geodlen{\ep_\mu,r_\mu}$. Hence, we use
$\geodlen{\ep_\nu,s_\nu} - \geodlen{\ep_\nu,r_\nu}$ as the \emph{event value}
of node $\nu$. This creates a strict ordering of the event values, and
thus of the event points in
the Voronoi cell of $p$. Unfortunately in many cases the first vertex
on the path towards $p$ or $q$ will not be the same for every vertex
on the bisector. Therefore we introduce an offset value to allow
comparing event points that have different first vertices on their
paths towards $p$ and $q$.

If $s_\nu \neq s_\mu$ and $r_\nu \neq r_\mu$, we should compare based
on a common vertex on the paths towards $p$ and $q$, which may be any
combination of $s_\nu$ or $s_\mu$ and $r_\nu$ or $r_\mu$. As these
cases are analogous, we consider the case where $s_\nu$ and $r_\mu$
are on the shortest paths towards $p$ and $q$ respectively for both
event points. (Intuitively $s_\nu$ and $r_\mu$ are further towards $p$
and $q$.)

Now the values we would like to compare are
\[
\geodlen{\ep_\nu,s_\nu} - \geodlen{\ep_\nu,r_\mu} > \geodlen{\ep_\mu,s_\nu} - \geodlen{\ep_\mu,r_\mu}.
\]

However these are not what the event values store.  With some
rewriting, we find that the above inequality holds if and only if
\[
\geodlen{\ep_\nu,s_\nu} - \geodlen{\ep_\nu,r_\nu} > \geodlen{\ep_\mu,s_\mu} - \geodlen{\ep_\mu,r_\mu} - \geodlen{s_\nu,s_\mu} - \geodlen{r_\nu,r_\mu}.
\]
We call $- \geodlen{s_\nu,s_\mu} - \geodlen{r_\nu,r_\mu}$ the
\emph{offset} of $\ep_\mu$ with respect to $\ep_\nu$. In the other
three cases we get a similar definition of offset. Now each node
will store the maximum event value in its subtree as follows.

For a leaf $\nu$ the maximum is its own event value,
i.e. $\geodlen{\ep_\nu,s_\nu} - \geodlen{\ep_\nu,r_\nu}$. For an
internal node, it is the maximum over its own event value and the
maximum values of its children with their offsets with respect to
$\nu$ added. The event point that realizes this maximum event value is
then the first event point in the subtree of $\nu$ that will be on the
bisector. Hence, the maximum event value the root can then be used to
determine the first time an $2,2$-event happens among the bisector
vertices stored in the tree.

Note that the above data structure stores only a constant number of
certificates directly involving $p$ or $q$, all of which are stored at
the root of the tree. Therefore, we can efficiently support changes in
the movement of $p$ and $q$. Let $w$ be a polygon vertex that is on
the shortest paths from $p$ to all the bisector vertices stored in the
tree (note that such a vertex exists, as the bisector vertices are all
part of a single non-visible piece of the bisector). When the shortest
path from $p$ to $w$ changes, we only have to update the offset stored
at the root of the tree. Recomputing this offset may take $O(\log m)$
time, and so does updating the $O(1)$ certificates of the root in the
global event queue.

Furthermore, we can support splitting the tree, and therefore this
invisible-piece of the bisector, at a vertex in $O(\log^2 m)$ time,
since a split affects $O(\log m)$ nodes in the balanced binary search
tree, and recomputing the offsets (and the corresponding maximum event
values) takes $O(\log m)$ time per node. Similarly, we can handle
joining two invisible-pieces of bisector in $O(\log^2 m)$ time as
well. This then also means we can insert or delete individual bisector
vertices in $O(\log^2 m)$ time.

\subparagraph{Handling Events} We now replace part (iii) of the naive
structure with the data structure described above. As argued, we are
still guaranteed to detect all events. However, we can now handle them
in $O(\log^2 m)$ time, rather than $O(m\log m)$ time. When $\SPM_p$
changes, as $p$ becomes collinear with the first two polygon vertices
$u$ and $v$ on a shortest path to $w$, we can now update the
certificates in each of the pieces efficiently. When the anatomy of
the bisector changes, this may involve joining or splitting two of the
invisible pieces. Thus, this takes $O(\log^2 m)$ time. Updating the
certificates within in each piece can also be done in $O(\log^2 m)$
time as argued before. Similarly, handling the other bisector events
can be done in $O(\log^2 m)$ time. We therefore obtain the following
theorem:

\begin{theorem}
  \label{thm:kds_bisector}
  Let $p$ and $q$ be two sites moving linearly inside a simple polygon
  $P$ with $m$ vertices. There is a KDS that maintains the bisector
  $B_{pq}$ that uses $O(m)$ space and processes at most $O(m^3)$
  events, each of which can be handled in $O(\log^2 m)$
  time. Additionally it can support movement changes of $p$ and $q$ in
  $O(\log m)$ time and splitting the bisector at any given vertex in
  $O(\log^2 m)$ time.
\end{theorem}

\section{A Voronoi Center}
\label{sec:center}

\begin{figure}[tbh]
  \centering
  \includegraphics[scale=0.95]{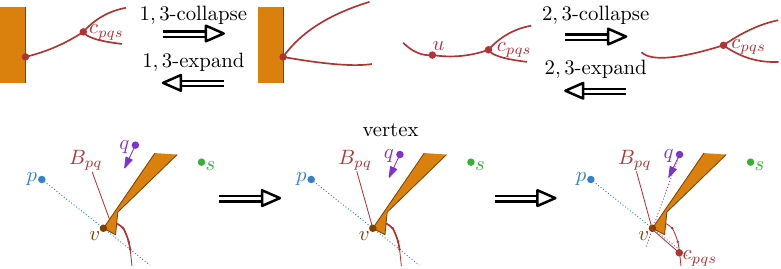}
  \caption{The events that can happen during the movement of a Voronoi center.}
  \label{fig:voronoi_center_events}
\end{figure}

Let $c_{pqs}(t)$ be the point equidistant to $p(t)$, $q(t)$, and
$s(t)$ if it exists. Note that Aronov~\etal~\cite{aronov1993furthest} proved that if it exists, it is unique (Lemma
2.5.3). We refer to $c_{pqs}$
as the \emph{Voronoi center} of $p$, $q$, and $s$. Note that there may
be times at which $c_{pqs}$ does not exist, and thus the trajectory of
$c_{pqs}$ may be disconnected. We identify five types of
events at which $c_{pqs}$ may appear or disappear, or at which the
movement of $c_{pqs}$ can change (see
Fig.~\ref{fig:voronoi_center_events}). They are:

\begin{itemize}[noitemsep]
\item \emph{$1,3$-collapse} events in which $c_{pqs}$ collides with
  the boundary of the polygon (in a bisector endpoint) and disappears
  from $P$,
\item \emph{$1,3$-expand} events in which $c_{pqs}$ appears on the
  boundary of $P$ as two bisector endpoints intersect, creating a
  point equidistant to all three sites,
\item \emph{vertex-events} where $c_{pqs}$ appears or disappears
  strictly inside $P$, as two sites, say $p$ and $q$, are equidistant
  to a vertex $v$ that appears on the shortest paths to $c_{pqs}$,
\item \emph{$2,3$-collapse} events where one of the geodesics
  from either $p$, $q$, or $s$ to $c_{pqs}$ loses a vertex,
\item \emph{$2,3$-expand} events where one of the shortest paths gains
  a new vertex.
\end{itemize}

Observe that, as the name suggests, at a $1,3$-collapse event the
Voronoi center (a degree 3 vertex in $\VD_P(\{p,q,s\})$) disappears as
it collides with the endpoint of a bisector (a degree 1
vertex). Similarly, at a $2,3$-collapse event a degree 2 vertex on one
of the bisectors disappears as it collides with a degree 3 vertex (the
Voronoi center $c_{pqs}$). As in case of the bisector, some of these
events may coincide. In the next section, we bound the number of
events, and thus the complexity of the trajectory of $c_{pqs}$. We
then present a kinetic data structure to maintain $c_{pqs}$ in
Section~\ref{sub:A_Kinetic_Data_Structure_center}.

\subsection{Bounding the Number of Events}
\label{sub:Bounding_the_Number_of_Events_center}

We give a construction in which the trajectory of $c_{pqs}$ has complexity
$\Omega(m^3)$, and then prove a matching upper bound.

\begin{lemma}
  \label{lem:lowerbound_center_pqs}
  The trajectory of the Voronoi center $c_{pqs}$ of three points $p$, $q$, and
  $s$, each moving linearly, may have complexity $\Omega(m^3)$.
\end{lemma}

\begin{proof}
 The main idea is that we can construct a trajectory for $c_{pqs}$ of
 complexity $\Omega(m^2)$, even when two of the three sites, say $p$ and $q$,
 are static. We place $p$ and $q$ so that their bisector $B_{pq}$, a piecewise
 hyperbolic curve of complexity $\Omega(m)$, intersects an (almost) horizontal
 line $E$ $\Omega(m)$ times. We can realize this using two convex chains $F_p$
 and $F_q$ in $\partial P$ similar to Lemma~\ref{lem:lowerbound_bisector}. See Fig.~\ref{fig:lowerbound_voronoi_center} for
 an illustration. We now construct a third convex chain $D_s$ in $\partial P$
 and place the third site $s$ so that the extension segments in $\SPM_s$
 incident to the vertices of $D_s$ all lie very close to $E$. Thus, each such
 segment intersects $B_{pq}$ $\Omega(m)$ times.  We choose the initial
 distances so that the voronoi center $c_{pqs}$ lies on the rightmost
 segment of $B_{pq}$. Now observe that as $s$ moves away from $D_s$, the
 center $c_{pqs}(t)$ will move to the left on $B_{pq}$, and thus it will pass
 over all $\Omega(m^2)$ intersection points of $B_{pq}$ with the extension
 segments of the vertices in $D_s$. At each such time, the structure of one
 of the shortest paths $\geod(p(t),c_{pqs}(t))$, $\geod(q(t),c_{pqs}(t))$, or
 $\geod(s(t),c_{pqs}(t))$ changes (they gain or lose a vertex from $F_p$,
 $F_q$, or $D_s$, respectively). Hence, the trajectory of $c_{pqs}$ changes
 $\Omega(m^2)$ times.

 Next, we argue that we can make $c_{pqs}$ ``swing'' from left to right
 $\Omega(m)$ times by having $p$ and $q$ move as well. The Voronoi center
 $c_{pqs}$ will then encounter every intersection point on $B_{pq}$
 $\Omega(m)$ times. It follows that the complexity of the trajectory of
 $c_{pqs}$ is $\Omega(m^3)$ as claimed.

 \begin{figure}[tbh]
   \centering
   \includegraphics[page=2,clip,trim=7cm 0.5cm 0 0]{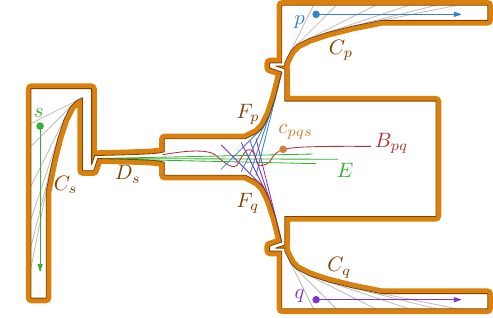}
   \caption{A
     polygon in which the trajectory of a voronoi center $c_{pqs}$ has
     complexity $\Omega(m^3)$.}
   \label{fig:lowerbound_voronoi_center}
 \end{figure}

 The idea is to add two additional convex chains, $C_s$ and $C_p$, that make
 the bisector $B_{ps}$ between $p$ and $s$ ``zigzag'' $\Omega(m) $ times
 throughout the movement of $p$ and $s$. We can achieve this using a similar
 construction as in Lemma~\ref{lem:lowerbound_bisector}. To make sure that the bisector
 $B_{pq}=B_{pq}(t)$ between $p$ and $q$ remains static, we create a third
 chain $C_q$, which is a mirrored copy of $C_p$, and we make $q$ move along a
 trajectory identical to that of $p$. See
 Fig.~\ref{fig:lowerbound_voronoi_center}. Finally, observe that
 $c_{pqs}(t) = B_{pq} \cap B_{ps}(t)$, and thus $c_{pqs}(t)$ will indeed
 encounter all $\Omega(m^2)$ intersection points on $B_{pq}$ $\Omega(m)$
 times. The lemma follows.
\end{proof}

\begin{lemma}
  \label{lem:13collapse_upperbound_center}
  The number of $1,3$-collapse events is at most $O(m^2)$.
\end{lemma}

\begin{proof}
  At a $1,3$-event $c_{pqs}$ to exits the polygon. Observe that at such a time
  $t$ a pair of bisector endpoints, say $b_{pq}(t)$ and $b_{ps}(t)$,
  intersect. By Lemma~\ref{lem:complexity_bpq} the trajectories of $b_{pq}$ and
  $b_{ps}$ have $O(m^2)$ edges, each of which is a constant degree algebraic
  curve. Thus, there are $O(m^2)$ time intervals during which both $b_{pq}$ and
  $b_{ps}$ move along the boundary of $P$, and their movement is described by a
  constant degree algebraic function. So such a time interval, $b_{pq}$ and $b_{ps}$
  coincide only $O(1)$ times. It follows that the total number of
  $1,3$-collapse events involving $p$, $q$, and $s$ is $O(m^2)$.
\end{proof}

\begin{lemma}
  \label{lem:center_vertex_events}
  The number of vertex events is at most $O(m^2)$.
\end{lemma}

\begin{proof}
 Fix a vertex $v$ and a pair of sites $p, q$. By
 Lemma~\ref{lem:closest_to_vertex} the site among $p,q$ closest to
 $v$ can change at most $O(m)$ times. Therefore, $v$ can produce at
 most $O(m)$ vertex events due to the pair $p,q$. Summing over all $m$
 vertices and all $O(1)$ pairs gives us an $O(m^2)$ bound.
\end{proof}

\begin{theorem}
  \label{thm:center_pqs}
  The trajectory of the Voronoi center $c_{pqs}$ has complexity $O(m^3)$. Each
  edge is a constant degree algebraic curve.
\end{theorem}

\begin{proof}
 A vertex in the trajectory of $c_{pqs}$ corresponds to either a
 $1,3$-collapse or expand, a vertex event, or a $2,3$-collapse or
 expand. By Lemma~\ref{lem:13collapse_upperbound_center} the number
 of $1,3$-collapse events, and symmetrically, the number of
 $1,3$-expand events, is $O(m^2)$. By
 Lemma~\ref{lem:center_vertex_events} the number of vertex events is
 also $O(m^2)$. We now bound the number of $2,3$-collapse (and
 symmetrically $2,3$-expand) events by $O(m^3)$. Each such an event
 corresponds to a breakpoint in the distance function between
 $c_{pqs}$ and one of the three sites. Hence, at such a time $t$,
 $c_{pqs}$ leaves an (extended) shortest path map cell $C_p$ in one
 of the three shortest path maps, say $\SPM_p$, and enters a
 neighboring cell of $\SPM_p$. Let $C_q$ and $C_s$ be the extended
 shortest path map cells in $\SPM_q$ and $\SPM_s$
 containing $c_{pqs}(t)$, respectively.

 All cells in the (extended) shortest path map of $p$ are triangles, and the
 map changes only $O(m)$ times throughout the movement of $p$
 (Lemma~\ref{lem:kinetic_spm}). Hence, $C_p$ corresponds to a constant complexity
 region $C'_p$ in $P \times \Time$ whose boundaries are formed by
 constant degree algebraic surfaces, and there are $O(m)$ such regions in
 total. Similarly, we have $O(m)$ choices for the constant complexity regions
 $C'_q$ and $C'_s$ corresponding to $C_q$ and $C_s$. Observe that within $C'_p
 \cap C'_q \cap C'_s$, all points have the same combinatorial shortest paths to
 $p$, $q$, and $s$, and thus the distance functions are continuous hyperbolic
 functions. Given these distance functions, the trajectory of $c_{pqs}$ is a
 constant degree algebraic curve. Such a curve can intersect the boundary of $C'_p
 \cap C'_q \cap C'_s$ at most $O(1)$ times. It follows that the maximum
 complexity of $c_{pqs}$ is thus $O(m^3)$.
\end{proof}

\subsection{A Kinetic Data Structure to Maintain a Voronoi Center}
\label{sub:A_Kinetic_Data_Structure_center}

\begin{figure}[tbh]
  \centering
  \includegraphics{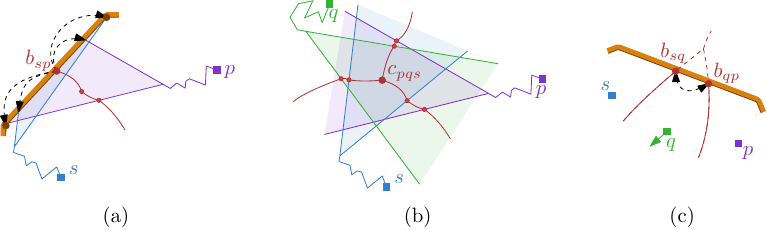}
  \caption{In black the certificates that we maintain in order to
    detect: (a) events where $b_{sp}$ changes movement, (b)
    $1,3$-collapse, $2,3$-collapse and $2,3$-expand events, and (c)
    $1,3$-expand events.}
  \label{fig:kds_certs}
\end{figure}

Our KDS for maintaining $c_{pqs}$ stores: (i) the extended shortest
path maps of $p$, $q$, and $s$, (ii) the cells of these shortest path
maps containing $c_{pqs}$ (when $c_{pqs}$ lies inside $P$), and (iii)
the endpoints of all bisectors (for all pairs), and their cyclic order
on $\partial P$. In particular, for each such endpoint $b_{sp}$ we
keep track of the cells of $\SPM_p$ and $\SPM_s$ that contain it. See
Fig.~\ref{fig:kds_certs} for an illustration. At any time we maintain
$O(m)$ certificates, which we store in a global priority queue.

Observe that at $1,3$-collapse, $2,3$-collapse, and $2,3$-expand
events the shortest path from $c_{pqs}$ to one of the sites changes
combinatorially. Hence, we can successfully detect all such events. At
a vertex event a vertex is equidistant to two sites, say $p$ and
$q$. At such a time, one of the two endpoints of $B_{pq}$ leaves an
edge of $P$, and thus exits a shortest path map cell in $\SPM_p$ (and
$\SPM_q$). Since we explicitly track all bisector endpoints, we can
thus detect this vertex event of $c_{pqs}$. Finally, at every
$1,3$-expand event two such bisector endpoints collide, and thus
change their cyclic order along $\partial P$. We detect such events
due to certificates of type (iii).

Any time at which $c_{pqs}$ changes cells in a shortest path map
results in a combinatorial change of its movement. Hence, any failure
of a certificate of type (ii) is an external event (a $1,3$-collapse,
$2,3$-collapse, or $2,3$-expand). The certificates of types (i) and
(iii) may be internal or external.

\begin{theorem}
  \label{thm:kds_center}
  Let $p, q$ and $s$ be three sites moving linearly inside a simple
  polygon $P$ with $m$ vertices. There is a KDS that maintains the
  Voronoi center $c_{pqs}$ that uses $O(m)$ space and processes at
  most $O(m^3)$ events, each of which can be handled in $O(\log^2 m)$
  time. Updates to the movement of $p$, $q$, and $s$, can be handled
  in $O(\log^2 m)$ time.
\end{theorem}

\begin{proof}
  Certificate failures of type (i) are handled exactly as described by
  Aronov \etal~\cite{agtz-vqmsp-02}. This takes $O(\log m)$
  time (by again implementing the dynamic convex hull queries using
  the Brodal and Jakob structure~\cite{brodal02dynam}). Note that changes to the shortest path maps may affect the
  certificates that guarantee that $c_{pqs}$ or a bisector endpoint
  lies in a particular SPM cell. In these cases we trigger a type (ii)
  or type (iii) certificate failure. At a certificate failure of type
  (ii) at which $c_{pqs}$ exits a shortest path map cell, we remove
  all certificates of type (ii) from the event queue. Next, for each
  site $p$, $q$, and $s$, we compute the new cell in the shortest path
  map containing $c_{pqs}$ (if $c_{pqs}$ still lies inside
  $P$). Finally, we create the appropriate new type (ii)
  certificates. Since all cells have constant complexity, the total
  number of certificates affected is also $O(1)$. Computing them can
  be done in $O(\log m)$ time, e.g. by re-querying the shortest path maps.

  Certificate failures of type (iii) where the movement of a bisector
  endpoint changes are handled using the same approach as in
  Section~\ref{sub:A_KDS_for_maintaining_the_Bisector}. Furthermore,
  at such an event we check if $c_{pqs}$ appears or disappears, that
  is, if the event is actually a vertex event of $c_{pqs}$. In this
  case, we recompute $c_{pqs}$ in $O(\log^2 m)$ time~\cite{oh_ahn2017voronoi_journal}. If
  $c_{pqs}$ disappears then we delete all type (ii) certificates. If
  $c_{pqs}$ appears then we locate the cell of $\SPM_p$, of $\SPM_q$,
  and of $\SPM_s$ that contains $c_{pqs}$, and insert new type (ii)
  certificates that certify this. Finding the cells and updating the
  certificates can be done in $O(\log m)$ time. At a certificate
  failure of type (iii) where two bisector endpoints collide, we check
  if the intersection point is equidistant to all three sites, and is
  thus a $1,3$-expand event. Similarly to the approach described
  above, we add new type (ii) certificates in $O(\log m)$ time in this case. 

  Maintaining the extended shortest path maps requires handling $O(m)$
  events~\cite{agtz-vqmsp-02}. Events where $c_{pqs}$ crosses a
  boundary of an extended SPM correspond to changes in the trajectory
  of $c_{pqs}$. By Theorem~\ref{thm:center_pqs} there are at most
  $O(m^3)$ such events. This dominates the $O(m^2)$ events that we
  have to handle to maintain the bisector endpoints in cyclic order
  around $\partial P$ (Lemmas~\ref{lem:complexity_bpq}
  and~\ref{lem:13collapse_upperbound_center}).

  In addition to $\SPM_p$, $\SPM_q$, and $\SPM_s$, we maintain
  only a constant amount of extra information. Since the KDS to
  maintain such a shortest path map $\SPM_s$ is local and supports changing the movement of $s$ in $O(\log^2 m)$ time, the
  same applies for our data structure as well. We therefore obtain a
  compact, responsive, local, and efficient KDS.
\end{proof}

\section{The Geodesic Voronoi Diagram}
\label{sec:Voronoi_diagram}

\begin{figure}[tbh]
  \centering
  \includegraphics[scale=0.75]{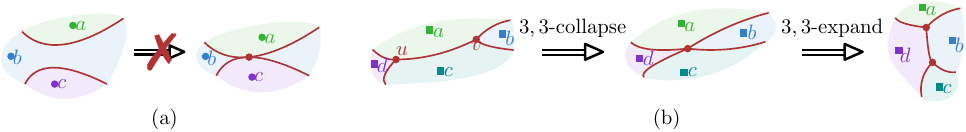}
  \caption{(a) Voronoi edges cannot intersect in their interior. (b) The
    $3,3$-collapse/expand events.}
  \label{fig:voronoi_events}
\end{figure}

In this section we consider maintaining the geodesic Voronoi diagram $\VD_P(S)$
as the sites in $S$ move. As a result of the sites in $S$ moving, the Voronoi
vertices and edges in $\VD_P(S)$ will also move. However, we observe that all
events involving Voronoi edges involve their endpoints; two edges cannot start
to intersect in their interior as this would split a Voronoi region, see
Fig.~\ref{fig:voronoi_events}(a). Similarly, the interior of a Voronoi edge
cannot start to intersect the polygon boundary. This means we can distinguish
the following types of events that change the combinatorial structure of the
Voronoi diagram.

\begin{itemize}[noitemsep]
\item Edge collapses, at which an edge between vertices $u$ and $v$ shrinks to
  length zero. Let $d_u,d_v$, with $d_u \leq d_v$, be the degrees of $u$ and
  $v$, respectively. We then have a \emph{$d_u,d_v$-collapse}.
\item Edge expands. These are symmetric to edge collapses.
\item Vertex events, where a degree 1 vertex of $\VD_P(S)$ crosses over a
  polygon vertex.
\end{itemize}

Indeed, we have seen most of these events when maintaining an
individual bisector or Voronoi center (a degree 3 vertex in
$\VD_P(S)$). The only new types of events are the $3,3$-collapse and
$3,3$-expand events which involve two degree 3 vertices. They are
depicted in Fig.~\ref{fig:voronoi_events}.(b). We again note that some
of these events may happen simultaneously.

\begin{theorem}
  \label{thm:complexity_bounds}
  Let $S$ be a set of $n$ sites moving linearly inside a simple
  polygon $P$ with $m$ vertices. During the movement of the sites in
  $S$, the combinatorial structure of the geodesic Voronoi diagram
  $\VD_P(S)$ changes at most $O(m^3n^3\beta_z(n))$ times. In
  particular, the events at which $\VD_P(S)$ changes, and the number
  of such events, are listed in Table~\ref{tab:voronoi_events}.
\end{theorem}

We prove these bounds in
Section~\ref{sub:Bounding_the_Number_of_Events_gvd}. For most of the
lower bounds we generalize the constructions from
Sections~\ref{sec:Bisector} and~\ref{sec:center}. For the upper bounds
we typically fix a site or vertex (or both), and map the remaining
sites to a set of functions in which we are interested in the lower
envelope. In Section~\ref{sub:A_KDS_for_a_Voronoi_Diagram} we develop
a kinetic data structure to maintain $\VD_P(S)$.

\subsection{Bounding the Number of Events}
\label{sub:Bounding_the_Number_of_Events_gvd}

We analyze the number of collapse events and the number of vertex events. The
expand events are symmetric to the collapse events.

\subsubsection{1,2-collapse Events}
\label{sub:1,2-collapse_Events}

\begin{lemma}
  \label{lem:lb_12_events}
  There may be $\Omega(m^2n)$ $1,2$-collapse events.
\end{lemma}

\begin{proof}
  \begin{figure}[tb]
    \centering
    \includegraphics{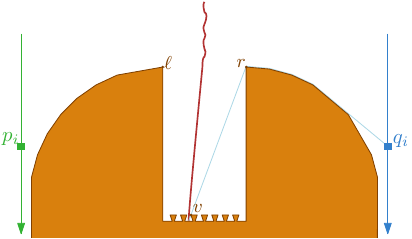}
    \caption{An illustration of the construction that yields $\Omega(m^2n)$
      $1,2$-collapse and also $\Omega(m^2n)$ vertex events. Note that
      the distances are not drawn exactly. }
    \label{fig:lb_vertex_events}
  \end{figure}
  We construct a polygon with a ``pit'' and two chains $D_p$ and $D_q$
  of reflex vertices, and we place $\Omega(m)$ ``$T$-shaped''
  obstacles at the bottom of the pit. See
  Fig.~\ref{fig:lb_vertex_events}. We make the sites move downwards in
  pairs $(p_i,q_i)$. All sites have the same speed, and they are
  spaced sufficiently far apart so that sites of different pairs do
  not interfere with each other. Site $p_i$ moves down left of the
  pit, so that for a sufficiently large time $t$ the shortest path
  from $p_i(t)$ to the ``$T$-shaped'' obstacles in the pit goes via
  $D_p$. Symmetrically, $q_i$ moves downwards on the right side of the
  pit. As in Lemma~\ref{lem:lowerbound_bisector} we make the chains
  $D_p$ and $D_q$ so that the bisector between $p_i$ and $q_i$ moves
  from left to right over the obstacles in the pit $\Omega(m)$ times.
  Now consider the ray from the left endpoint $r$ of the top right
  chain $D_q$ through the top-left vertex $v$ of a $T$-shaped
  obstacle. Let $a$ be the point where this ray hits the floor of the
  pit. The site closest to $a$ changes $\Omega(m)$ times from $p_i$ to
  $q_i$ (as the endpoint of the bisector sweeps from right to
  left). Every such a time corresponds to a $1,2$-collapse
  event. Since we have $\Omega(n)$ pairs, the lemma follows.
\end{proof}

\begin{lemma}
  \label{lem:12collapse}
  The number of $1,2$-collapse events is at most $O(m^2n^2)$.
\end{lemma}

\begin{proof}
  Any $1,2$-collapse event in the Voronoi diagram uniquely corresponds to a
  $1,2$-collapse event of a bisector $B_{pq}(t)$ for some sites $p,q \in S$. By
  Lemma~\ref{lem:12_collapse_pq} the number of $1,2$-collapse events in
  $B_{pq}(t)$ is $O(m^2)$. The lemma follows by summing over all pairs
  $p,q \in S$.
\end{proof}

\subsubsection{1,3-collapse Events}
\label{sub:1,3-collapse_Events}

\begin{figure}[tbh]
  \centering
  \includegraphics{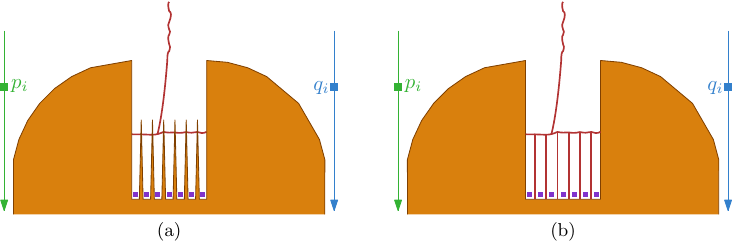}
  \caption{The construction that yields
    $\Omega(nm\min\{n,m\})$ $1,3$-collapse events (a), and $\Omega(mn^2)$
    $3,3$-collapse events (b). The distances are again not drawn exactly. }
  \label{fig:lb_voronoi_flips}
\end{figure}

\begin{lemma}
  \label{lem:13collapse_lowerbound}
  There may be $\Omega(mn\min\{n,m\})$ $1,3$-collapse events.
\end{lemma}

\begin{proof}
  We use a similar construction with a ``pit'' and $\Omega(n)$ pairs
  of entities as in
  Lemma~\ref{lem:lb_12_events}. See Fig.~\ref{fig:lb_voronoi_flips}(a). Place $\Omega(\min\{n,m\})$ spikes at the bottom of the pit and place a site between them. As two sites $p_i$ and $q_i$ move down, their bisector sweeps over all spikes causing the voronoi center of $p_i$, $q_i$ and the site between the spikes as the bisector to hit the side of each spike, causing a $1,3$-collapse event each time. Since the upper convex chains of the polygon consist of $\Omega(m)$ vertices, every pair of vertices causes $\Omega(m \min\{n,m\})$ such events. Note that while the horizontal part of the bisectors between the spikes moves up as a pair of vertices moves down, it ``resets'' when a new pair arrives, thus this process can be repeated $\Omega(n)$ times using new sites each time and the lemma follows.
\end{proof}

By Lemma~\ref{lem:13collapse_upperbound_center} any triple $p$, $q$, $s$
generates at most $O(m^2)$ $1,3$-collapse events. So, summing over all triples
this immediately gives us an $O(m^2n^3)$ upper bound. Next, we argue that we
can also bound the number of $1,3$-collapses by $O(m^3n^2\beta_z(n))$, for
some $z \in \mathbb{N}$.

\begin{lemma}
  \label{lem:13collapse_upperbound2}
  The number of $1,3$-collapse events is at most $O(m^3n^2\beta_z(n))$.
\end{lemma}

\begin{proof}
  Fix a site $s$ and an edge $e$ of the polygon. We now bound the number of
  $1,3$-collapse events on site $e$ involving site $s$ by $O(m^2n\beta_z(n))$,
  for some constant $z$. Since we have $n$ sites and $m$ edges, the lemma then
  follows. At any time $t$, the Voronoi region of $s$ intersects $e$ in at
  most a single connected interval~\cite{dyn_geod_nn2018}. 
  All $1,3$-events on $e$ involving $s$ occur on one of the two endpoints of this
  interval. Let $a(t)$ be the endpoint such that $s$ lies right of the edge of
  $\VD(t)$ that starts in $a(t)$. See
  Fig.~\ref{fig:13_collapse_upperbound}. Next, we bound the number of events
  occurring at $a(t)$. Bounding the number of events occurring at the other endpoints is analogous.
  \begin{figure}[tbh]
    \centering
    \includegraphics{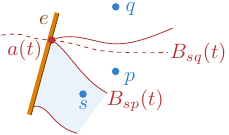}
    \caption{At $1,3$-collapse event two bisector endpoints $b_{sp}(t)$ and $b_{sq}(t)$ meet.}
    \label{fig:13_collapse_upperbound}
  \end{figure}
  Observe that $a(t)$ is a bisector endpoint $b_{sp}(t)$ for some site
  $p$. More specifically, it is the ``lowest'' such endpoint along $e$. At an
  $1,3$-collapse event, the site that defines this ``lowest'' endpoint changes,
  that is, two bisector endpoints $b_{sp}(t)$ and $b_{sq}(t)$ meet. More
  formally, let $e=\overline{uv}$ and let $\lambda(w) \in [0,1]$ be the value
  such that $w=((1-\lambda(w))u + \lambda(w)v)$ for all points $w \in e$. For a
  site $r$ we then define the function
  \[ f_r(t) =
    \begin{cases}
      \lambda(b_{sr}(t)) & \text{if } b_{sr}(t) \text{ lies on }  e \\
      \infty             & \text{otherwise}.
    \end{cases}
  \]
  If now follows that a $1,3$-collapse event at $a(t)$ corresponds to a
  vertex in the lower envelope $\Lst(\{f_r \mid r \in (S\setminus
  \{s\})\})$. Since the trajectory of any $b_{sr}$ has complexity $O(m^2)$
  whose edges are low-degree algebraic curves (Lemma~\ref{lem:complexity_bpq})
  that pairwise intersect $O(1)$ times, the same applies for function $f_r$. It
  follows that their lower envelope has complexity $O(m^2n\beta_z(n))$, for
  some constant $z$~\cite{sharir1995davenport}, and thus the number of
  $1,3$-collapses at $u(t)$ is at  most $O(m^2n\beta_z(n))$ as well.
\end{proof}

\begin{corollary}
  \label{cor:13collapse_upperbound}
  The number of $1,3$-collapse events is at most\\
  $O(m^2n^2\min\{m\beta_z(n),n\})$.
\end{corollary}

\subsubsection{2,2-collapse Events}
\label{sub:2,2-collapse_Events}

\begin{lemma}
  \label{lem:22collapse_lowerbound}
  There may be $\Omega(m^3n)$ $2,2$-collapse events.
\end{lemma}

\begin{proof}
  We use the construction from Lemma~\ref{lem:lowerbound_bisector} in which the
  bisector of a single pair of sites $(p,q)$ changes $\Omega(m^3)$ times. We
  now simply create $\Omega(n)$ such pairs $(p_i,q_i)$ that all move along the
  same trajectories. We choose the starting positions such that the distance
  between two consecutive points $p_i$ and $p_{i+1}$ ($q_i$ and $q_{i+1}$) is
  very large, so that for each pair $(p_i,q_i)$ the bisector appears in the
  Voronoi diagram at the time when $p_i$ and $q_i$ pass by our construction.
\end{proof}

\begin{lemma}
  \label{lem:22collapse_upperbound}
  The number of $2,2$-collapse events is at most $O(m^3n\beta_6(n))$.
\end{lemma}

\begin{proof}
  Fix two vertices $u$ and $v$. By Lemma~\ref{lem:closest_to_vertex} there are
  a total of $O(mn\beta_6(n))$ maximal time intervals during which both $u$ and
  $v$ have unique closest sites, and the distances from $u$ and $v$ to their
  respective closest sites, say $r$ and $s$, is a continuous hyperbolic
  function. Consider such an interval $I$, and observe that during $I$, there
  is only a single extension segment $E_{ur}$ in the (non extended) $\SPM_r$,
  and thus in $\VD$, incident to $u$. Similarly, there is a single extension
  segment $E_{vs}$ in $\VD$ incident to $v$. Like in
  Lemma~\ref{lem:22collapse_upperbound_pq}, we now have that in $I$ the
  distances from $r$ and $s$ to the intersection point of $E_{ur}$ and $E_{vs}$
  are both continuous algebraic functions of constant degree. These functions intersect
  only a constant number of times, and thus there are at most $O(1)$
  $2,2$-collapse events per interval. In total we thus have $O(mn\beta_6(n))$
  events per pair, and $O(m^3n\beta_6(n))$ events in total.
\end{proof}

\subsubsection{2,3-collapse Events}
\label{sub:2,3-collapse_Events}

\begin{lemma}
  \label{lem:23collapse_lowerbound}
  There may be $\Omega(mn^2 + m^3n)$ $2,3$-collapse-events.
\end{lemma}

\begin{proof}
  We give two constructions. The first one gives $\Omega(m^3n)$ $2,3$-collapse
  events and the second one $\Omega(mn^2)$. The lemma then follows.

  For the first construction we modify the construction in
  Fig.~\ref{fig:lowerbound_voronoi_center} slightly. The main idea is
  that each time the Voronoi center moves into a new cell defined by
  $F_p$, $F_q$, and $D_s$ a $2,3$-collapse event occurs. Thus for
  three sites, we get $\Omega(m^3)$ such events. When we repeat this
  process by moving $\Omega(n)$ triples along the trajectories of $p$,
  $q$, and $s$, the first part of the lower bound follows. In order to
  do this, we modify the polygon by adding two horizontal rectangles,
  one to the left of $p$ and one to the left of $q$, and a vertical
  rectangle, above $s$. These rectangles contain (the trajectories of)
  the future triples. By making the convex chains $C_p$, $C_q$, and
  $C_s$ steep enough, we can ensure that all events occur close enough
  together, limiting how long the horizontal rectangles need to
  be. Thereby, we can make sure they do not overlap the vertical path
  of $s$. Alternatively, we can make the part of the polygon containing
  $s$ horizontal, and use the ``zigzags'' between $C_p$ and $F_p$ and
  between $C_q$ and $F_q$ to adapt the length of the paths appropriately.

    \begin{figure}[tb]
    \centering
    \includegraphics{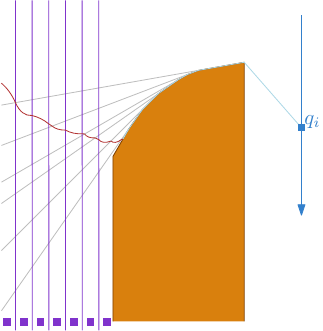}
    \caption{An illustration of the construction that yields
      $\Omega(mn^2)$ $2,3$-collapse events.}
    \label{fig:lb_23collapse2}
  \end{figure}
  The second construction is sketched in Fig.~\ref{fig:lb_23collapse2}. We again have an obstacle with a convex chain of complexity $\Omega(m)$. On the left side of this obstacle, we have $\Omega(n)$ fixed sites. The bisectors of adjacent sites and the lines extending the edges of the convex chain form a grid of complexity $\Omega(mn)$. On the right side of the obstacle, we drop a site $q_i$ such that the bisector of $q_i$ and the fixed sites sweeps over the entire grid, causing $\Omega(nm)$ 2,3-collapse events. By dropping $\Omega(n)$ sites sufficiently far apart, we can repeat this process $\Omega(n)$ times, leading to the second part of the lower bound.
\end{proof}

Consider an extension segment $E_{vp}(t)$ of $\SPM_p(t)$ incident to $v$ and
let $\lambda \in [0,1]$ be some linear parameter along $E_{vp}(t)$ such that
$E_{vp}(t,\lambda) = (1-\lambda)v + \lambda{}e_{vp}(t)$ is a point along
$E_{vp}(t)$. Let $f_{vp,q}(t,\lambda)=\geodlen{q(t),E_{vp}(t,\lambda)}$ denote the
distance function from a site $q(t)$ to $E_{vp}(t,\lambda)$.

\begin{lemma}
  \label{lem:distance_function_along_extension_seg_single}
  Consider a time interval in which $\SPM_q$ has fixed combinatorial
  structure. The function $f_{vp,q}$ restricted to this interval is a bivariate
  piecewise constant degree algebraic function of complexity $O(m)$.
\end{lemma}

\begin{proof}
  Every cell of $\SPM_q$ has constant complexity. Moreover, the function
  describing the movement of $E$ has constant complexity as well. In
  particular, this function consists of at most three pieces, in at most one of
  which $E$ lies on the (rotating) line through $p$ and $v$, and in the other
  two $E$ has a fixed location. It follows that each $\SPM_q$ cell contributes
  constant complexity to $f_{vp,q}$. Since there are only $O(m)$ cells, the
  total complexity of $f_{vp,q}$ is at most $O(m)$. Every patch of $f_{vp,q}$
  represents the Euclidean distance of a fixed point (a polygon vertex) to a
  points on a line rotating through $v$. This can be described by a
  constant degree
  algebraic function.
\end{proof}

\begin{lemma}
  \label{lem:23collapse_upperbound}
  The number of $2,3$-collapse events is at most
  $O(m^3n^2\beta_6(n)\beta_z(n))$.
\end{lemma}

\begin{proof}
  Any $2,3$-collapse event at some time $t$ occurs on an extension segment
  $E(t)$ incident to a polygon vertex $v$. In particular, $E(t)$ is an
  extension segment of the shortest path map of the site $s_v(t)$ which is
  closest to $v$ at time $t$. Furthermore, by Lemma~\ref{lem:structure_gvd}, $v$
  has only one such an extension segment at any time. We can thus charge the
  $2,3$-event to $v$. We now bound the number of such charges to a vertex $v$
  by $O(m^2n^2\beta_6(n))$. The lemma then follows.

  Split time into time intervals in which: (\emph{i}) the site $s_v(t)$ closest
  to $v$ is fixed, and the distance from $s_v(t)$ to $v$ is a continuous hyperbolic
  function, and (\emph{ii}) the shortest path maps from all sites have a fixed
  combinatorial structure. It follows from Lemmas~\ref{lem:kinetic_spm} and
  ~\ref{lem:closest_to_vertex} that there are $O(mn\beta_6(n))$ intervals in
  total.

  \begin{figure}[tb]
    \centering
    \includegraphics{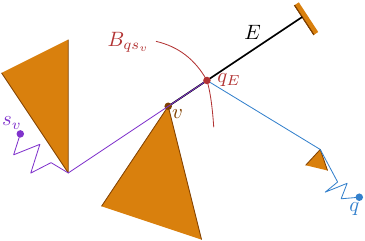}
    \caption{The extension segment $E$ incident to $v$ defined by the site
      $s_v$ closest to $v$, and the intersection point $q_E$ between $E$ and
      the bisector of $q$ and $s_v$. The function $g_q$ measures the distance
      from $q_E$ to $s_v$, that is, the length of the red and blue paths.}
    \label{fig:23collapse_defs}
  \end{figure}

  Fix such a time interval $I$. For any (other) site $q \neq s_v$, let
  $q_E(t)$ be the intersection point of $E(t)$ and the bisector of $q$ and
  $s_v(t)$. See Fig.~\ref{fig:23collapse_defs}. The distance function $f_{E,q}$
  restricted to $I$ has complexity $O(m)$
  (Lemma~\ref{lem:distance_function_along_extension_seg_single}), and the
  distance from $s_v(t)$ to $E(t)$ has constant complexity. It follows that the
  complexity of the trajectory of $q_E(t)$ in the interval $I$ is also
  $O(m)$. In turn, this implies that the function
  $g_q(t)=\geodlen{s_v(t),q_E(t)}$ has only $O(m)$ breakpoints in interval
  $I$. Moreover, each piece of $g_q$ is some constant degree algebraic function. For
  $q=s_v$ and $t \in I$ we let $g_q(t)$ be undefined.

  Since we have $O(mn\beta_6(n))$ intervals, it follows that the function $g_q$
  has a total complexity of $O(m^2n\beta_6(n))$.

  Any $2,3$-collapse charged to $v$ now corresponds to a vertex on the lower
  envelope of the functions $g_q(t)$: at such a vertex two sites, say $q$ and
  $r$ are both closest to $s_v(t)$, and there is no other site closer. The
  lower envelope has complexity $O(m^2n\beta_6(n)n\beta_z(n))$. It follows that
  there are thus also at most $O(m^2n^2\beta_6(n)\beta_z(n))$ $2,3$-collapse
  events charged to $v$. The lemma follows.
\end{proof}

\subsubsection{3,3-collapse Events}
\label{sub:3,3-collapse_Events}

\begin{lemma}
  \label{lem:33collapse_lowerbound}
  There may be $\Omega(mn^2)$ $3,3$-collapse events.
\end{lemma}

\begin{proof}
  See Fig.~\ref{fig:lb_voronoi_flips}(b). As in
  Lemma~\ref{lem:13collapse_lowerbound}, we again we drop points in pairs, but now the
  bottom of the pit contains $\Omega(n)$ collinear sites $r_1,..,r_h$. As the bisector
  between $p_i$ and $q_i$ moves from left to right, it crosses the vertical
  bisector between $r_j$ and $r_{j+1}$, collapsing an edge $(u,v)$ of the
  Voronoi diagram. Since both $u$ and $v$ are degree 3 vertices, this is a
  $3,3$-collapse. It follows that every pair of sites $p_i$, $q_i$, generates
  $\Omega(mn)$ such events. The lemma follows.
\end{proof}

\input{33flip_quadratic_m_lowerbound}

Next, we prove an upper bound for the number of $3,3$-collapse
events. We use the same general idea as Guibas
\etal~\cite{guibas1991kineticvd} use for sites moving in $\R^2$ under
semi-algebraic motion.

Consider the bisector $B_{pq}$ of a pair of points $p,q$, and orient
it so that a point $c_s$ on $B_{pq}$ is above a point $c_r$ on the
bisector if the pseudo-triangle defined by $p,c_s,c_r$ has those
points in that clockwise order on its boundary, see
Fig.~\ref{fig:r_in_Ds}.

\begin{lemma}
  \label{lem:contained}
  Let $D_s$ be a geodesic disk containing $p$, $s$, and $q$ on its
  boundary in that order, let $D_r$ be a disk with $p$, $r$ and $q$ in
  that order on the boundary. The center $c_r$ of $D_r$ lies ``below''
  the center $c_s$ of $D_s$ in the order along $B_{pq}$ if and only if
  $D_s$ contains $r$.
\end{lemma}

\begin{proof}
  To show that $r$ lies inside $D_s$, we argue that
  $|\geod(c_s, r)| \leq |\geod(c_s, p)|$. Consider the geodesic
  triangle $\Delta$ defined by $p$, $c_s$, and $q$. The distance
  function from $c_s$ to points on $\geod(p,q)$ is
  convex~\cite{pollack89comput_geodes_center_simpl_polyg}, and thus
  has its maximum at $p$ or $q$. So, if $r$ lies inside $\Delta$, it
  now readily follows that
  $|\geod(c_s,r)| \leq |\geod(c_s,p)| = |\geod(c_s,q)|$.
  \begin{figure}[tbh]
    \centering
    \includegraphics{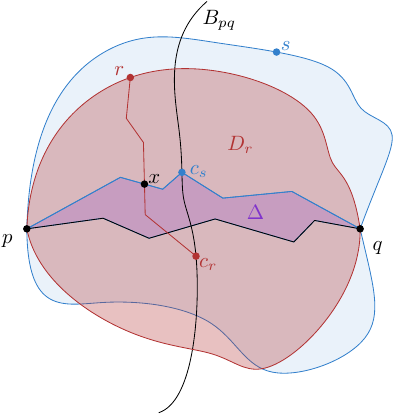}
    \caption{If $c_r$ lies ``below'' $c_s$ along the bisector $B_{pq}$
      and $r$ does not lie in the geodesic triangle defined by $p$,
      $c_s$, and $q$ then $\geod(c_r,r)$ must intersect either
      $\geod(c_s,p)$ or $\geod(c_s,q)$ in a point $x$. This allows us
      to argue that $|\geod(c_s,r)|\leq |\geod(c_s,p)|$; meaning that
      $r$ lies in $D_s$.  }
    \label{fig:r_in_Ds}
  \end{figure}
  If $r$ lies outside $\Delta$ then we claim that the shortest path
  $\geod(c_r,r)$ must intersect $\geod(p,c_s) \cup \geod(c_s, q)$. The
  geodesic triangle $\Delta$ splits $D_r$ into three pieces: $\Delta$
  itself, a part ``above'' $\Delta$ containing the points $z$ in $D_r$
  so that the geodesic triangle $p,z,q$ is oriented clockwise, and the
  remaining part ``below'' the $\Delta$. See Figure~\ref{fig:r_in_Ds}
  for an illustration. Since $r$ lies on the boundary of $D_r$ in
  between $p$ and $q$ (with respect to a clockwise order), and it does
  not lie inside $\Delta$, it must lie above
  $\geod(p,c_s) \cup \geod(c_s, q)$. Since $c_r$ lies ``below'' $c_s$
  (in the order along $B_{pq}$), it must lie below or inside the
  triangle. Finally, since $\geod(c_r,r)$ is contained in $D_r$, it
  must thus intersect $\geod(p,c_s) \cup \geod(c_s, q)$.

  Assume, without loss of generality that $\geod(c_r,r)$ intersects
  $\geod(p,c_s)$ in a point $x$. We then obtain by triangle inequality
  that:
  \begin{align*}
    |\geod(c_s, p)| &= |\geod(c_s, x)| + |\geod(x, p)| \\
    |\geod(c_s, r)| &\leq |\geod(c_s, x)| + |\geod(x, r)| \\
    |\geod(c_r, p)| &\leq |\geod(c_r, x)| + |\geod(x, p)| \\
    |\geod(c_r, r)| &= |\geod(c_r, x)| + |\geod(x, r)| 
  \end{align*}
  
  We thus obtain that 
  \begin{align*}
    |\geod(c_r, x)| + |\geod(x, r)| &= |\geod(c_r, r)| \\
    &= |\geod(c_r, p)| \\
    &\leq |\geod(c_r, x)| + |\geod(x, p)|,
  \end{align*}
  where the second equality follows from the fact that $p$ and $r$ are on the boundary of $D_r$. 
  This implies that $|\geod(x, r)| \leq |\geod(x, p)|$, which in turn gives us that
  \begin{align*}
    |\geod(c_s, r)| &\leq |\geod(c_s, x)| + |\geod(x, r)| \\
    &\leq |\geod(c_s, x)| + |\geod(x, p)| \\
    &= |\geod(c_s, p)|
  \end{align*}
  
  Hence, we obtain that $|\geod(c_s, r)| \leq |\geod(c_s, p)|$ and
  thus $r$ lies inside $D_s$.
\end{proof}

\begin{lemma}
  \label{lem:33collapse_upperbound}
  There are $O(m^3n^3\beta_z(n))$ $3,3$-collapse events.
\end{lemma}

\begin{proof}
  The main idea is as follows. Recall that a degree three Voronoi vertex
  $c_{pqs}(t)$ is defined by three sites, $p(t),q(t),s(t)$ that lie on
  the boundary of a geodesic disk $D_{pqs}(t)$ that is empty of other
  sites. At a $3,3$-collapse event at time $t$ two degree three Voronoi
  vertices $c_{pqs}$ and $c_{pqr}$ collide, and hence their empty
  geodesic disks $D_{pqs}(t)=D_{pqr}(t)$ coincide. Relabel the sites so
  that at the time of the event, the clockwise order of the points along
  the boundary of this disk $D_{pqs}(t)=D_{pqr}(t)$ is
  $p(t),r(t),s(t),q(t)$. We will charge the collapse event to the pair
  $p,q$, and argue that each such pair is charged at most
  $O(m^3n \beta_z(n))$ times. Therefore the total number of
  $3,3$-collapse events is at most $O(m^3n^3 \beta_z(n))$ as claimed.

  To bound the number of times each pair $p,q$ is charged, we map each
  other site $s$ to a function $\mu^+_{pqs}$, and argue that
  $3,3$-collapse events charged to $p,q$ correspond to vertices of the
  lower envelope of these functions. What remains is to describe these
  functions.

  \begin{figure}[tbh]
    \centering
    \includegraphics{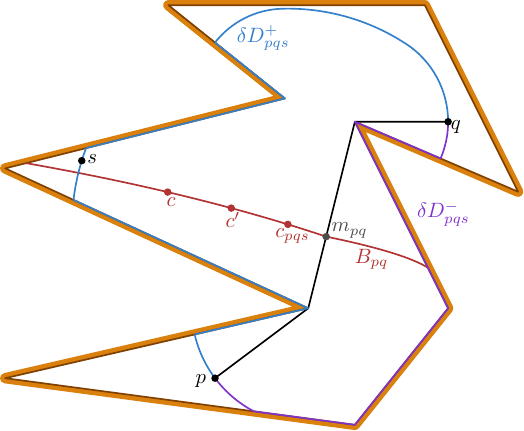}
    \caption{An illustration of the definitions used in the proof of
      Lemma~\ref{lem:33collapse_upperbound}.}
    \label{fig:upperbound_33collapse}
  \end{figure}

  For any site $s$, consider the geodesic disk $D_{pqs}(t)$ with
  $p(t)$, $q(t)$, and $s(t)$ on its boundary that has $c_{pqs}(t)$ as
  its centerpoint. At any time there is at most one such a
  disk~\cite{oh2016_2center,oh2018_2center}. Observe that (if the
  center point lies inside $P$) it lies on the bisector $B_{pq}(t)$,
  and that the points $p,q$ divide the boundary of $D_{pqs}(t)$ into
  two parts. Let $\delta D_{pqs}^+(t)$ denote the boundary section
  which is clockwise adjacent to $p$ and $\delta D_{pqs}^-(t)$ the
  part that is counterclockwise adjacent to $p$, see
  Fig.~\ref{fig:upperbound_33collapse}. The site $s$ can be on either
  boundary part. Let $S_{pq}^+(t)$ be the set of sites $s$, so that
  $c_{pqs}(t)$ is on $B_{pq}(t)$ (it may not be inside $P$ for all
  sites $s$) and $s(t)$ is on $\delta D_{pqs}^+(t)$.  Now observe that
  any $3,3$-collapse event $p,r,s,q$ at time $t$ charged to $p,q$ will
  have $s,r \in S_{pq}^+(t)$.

  Consider two sites $s,s' \in S^+(t)$, and their centers $c_{pqs}(t)$
  and $c_{pqs'}(t)$. By Lemma~\ref{lem:contained}, point $c_{pqs'}(t)$
  lies below $c_{pqs}(t)$ on the bisector $B_{pq}(t)$ if and only if
  the disk $D_{pqs}(t)$ contains $s'(t)$.

  At a $3,3$-event $p,r,s,q$ charged to $p,q$, the disks $D_{pqs}(t)$
  and $D_{pqr}(t)$ are both empty of other sites. Hence, it follows
  that their centers $c_{pqs}(t)$ and $c_{pqr}(t)$ are the ``lowest''
  centers among all sites in $S^+(t)$. Our final step is to formally
  capture this notion of lowest definition of $\mu^+(t)$, and argue
  about its complexity.

  Let $m_{pq}(t)$ denote the midpoint of the shortest path between
  $p(t)$ and $q(t)$ (and observe that $m_{pq}(t)$ lies on
  $B_{pq}(t)$). Then for a point $c$ on the bisector $B_{pq}(t)$ that
  is above $m_{pq}(t)$, we define
  $F_{pq}(t,c) = \dist(p(t),c) - \dist(p(t), m_{pq}(t))$. For a point
  $c''$ below $m_{pq}(t)$, we define
  $F_{pq}(t,c'') = - (\dist(p(t),c'') - \dist(p(t), m_{pq}(t)))$. Then
  for any site $s(t)$ not equal to $p(t)$ or $q(t)$, we define
  \[
    \mu_{pqs}^+(t) =
    \begin{cases}
      F_{pq}(t,c_{pqs}(t)) & \text{if } s(t) \in S_{pq}^+(t), \text{
        and }\\
      \infty &\text{otherwise.}
    \end{cases}
  \]
  For any point $s$, there are $O(m^3)$ time intervals in which the
  movement of the Voronoi center $c_{pqs}$, and the (lengths of the)
  shortest paths between the sites $p$, $q$, and $s$ defining
  $c_{pqs}$ are described by a constant degree algebraic
  function~(Theorem~\ref{thm:center_pqs}). For each such a time
  interval $s$ can cross $\geod(p,q)(t)$ and thus enter or leave $S^+$
  at most $O(1)$ times. Hence, the complexity of $\mu_{pqs}^+$ is also
  $O(m^3)$. Since we are considering $O(n)$ sites we are interested in
  the lower envelope of $O(n)$ functions, each consisting of $O(m^3)$
  pieces of constant algebraic complexity. Therefore the number of
  changes in the lower envelope, and thus the number of events charged
  to $p,q$ is bounded by
  $O(m^3n \beta_z(n))$~\cite{sharir1995davenport}. This completes the
  proof.
\end{proof}

\subsubsection{Vertex Events}
\label{sub:Vertex_Events}

\begin{lemma}
  \label{lem:vertex_events_lowerbound}
  There may be $\Omega(m^2n)$ vertex events.
\end{lemma}

\begin{proof}
  We once more use a similar approach as in
  Lemma~\ref{lem:13collapse_lowerbound}. We build the ``pit''
  construction shown in Fig.~\ref{fig:lb_vertex_events} and we drop
  the sites in pairs of two, say pairs $p_i,q_i$. The left and right
  convex chains have complexity $\Omega(m)$ and are built such that
  the endpoint of the bisector of $p_i$ and $q_i$ sweeps the
  $\Omega(m)$ ``T-shaped'' obstacles every time its geodesic path to
  $p_i$ or $q_i$ changes. It now follows that the point where the
  bisector of $p_i$ and $q_i$ hits the bottom of the pit moves across
  the obstacles in the pit $\Omega(m)$ times. Thus the number of
  vertex events is $\Omega(m^2n)$.
\end{proof}

\begin{lemma}
  \label{lem:vertex_events_upperbound}
  The number of vertex events is at most $O(m^2n\beta_6(n))$.
\end{lemma}

\begin{proof}
  Directly from Lemma~\ref{lem:closest_to_vertex}, summing over all vertices.
\end{proof}

\subsection{A KDS for a Voronoi Diagram}
\label{sub:A_KDS_for_a_Voronoi_Diagram}

\begin{figure}[tbh]
\centering
\includegraphics{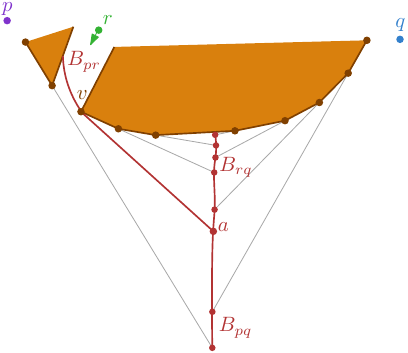}
\caption{A vertex event may split a bisector or a degree 3 vertex crossing an SPM extension segment may cause two bisectors to merge.}
\label{fig:bisector_split}
\end{figure}

In this section we develop a KDS to maintain the Voronoi diagram of
$S$. Our KDS essentially stores for each site the extended shortest
path map of its Voronoi cell, and a collection of certificates that
together guarantee that the shortest paths from the sites to all
Voronoi vertices remain the same (and thus the KDS correctly
represents $\VD_P(S)$). The main difficulties that we need to deal
with are shown in Fig.~\ref{fig:bisector_split}. Here, $r$ becomes the
site closest to vertex $v$, and as a result a part of the polygon
moves from the Voronoi cell $V_p$ of $p$ to the Voronoi cell $V_r$ of
$r$. Our KDS should therefore support transplanting this region from
the SPM representation of $V_p$ into $V_r$ and vice versa. Moreover,
part of the bisector $B_{pq}$ becomes a bisector $B_{pr}$, which means
that any certificates internal to the bisector (such as those needed
to detect 2,2-events) change from being dependent on the movement of
$p$ to being dependent on the movement of $r$. Next, we show how to
solve the first problem, transplanting part of the shortest path
map. Our KDS for the bisector from Theorem~\ref{thm:kds_bisector}
essentially solves the second problem. All that then remains is to
describe how to handle each event.

\paragraph{Maintaining Partial Shortest Path Maps}
To support transplanting a part of $\SPM_s$ into $\SPM_q$ we extend
the data structure of Aronov~et~al.~\cite{agtz-vqmsp-02}. Observe that
$\SPM_s$ is a tree rooted at $s$, and we transplant only subtrees,
rooted at some polygon vertex $v$. Our representation of $\SPM_s$
should support: (i) link operations in which we add the subtree rooted
at $v$ as a child of $u$, (ii) cut operations in which we cut an edge
$(u,v)$, (iii) shortest path queries in which we report the length of
the shortest path from some vertex $u$ to the root $s$, and (iv)
principal-child queries in which we report the \emph{principal child}
$c$ of some non-root node $u$. The principal child is the child of $u$
for which the angle between $\overline{cu}$ and $\overline{up(u)}$,
where $p(u)$ is the parent of $u$, is minimal. We need this operation
to support updating the certificates of $\SPM_s$\footnote{Since the
  root is the only node storing a moving point, all certificates
  involve only nodes from the first three layers of the tree. Hence,
  it suffices to compute the principal child only for direct children
  of the root.}. To support these operations, we store $\SPM_s$ twice:
once in a link-cut tree~\cite{sleator1983link_cut} and once in an
Euler tour tree~\cite{henzinger1999eulertour}. Both these structures
support link and cut operations in $O(\log m)$ time. The link-cut
trees support query operations on node-to-root paths, and hence we use
them to answer shortest path queries in $O(\log m)$ time (plus $O(k)$
time to report the actual path, if desired). The Euler tour trees
support query operations on subtrees, and hence we use them to answer
principal child queries. In particular, we maintain the children of
$u$ in cyclic order around $u$, starting with $c$. This way link and
cut operations still take $O(\log m)$ time, and the principal child of
$u$ can be reported in constant time.

\paragraph{The data structure}
The full KDS thus consists of an extended shortest path map for every
Voronoi cell maintained as described above; and certificates for each
degree three vertex, degree one vertex, and each bisector. For every
degree three vertex $c_{pqs}$ we maintain the cells of $\SPM_p$,
$\SPM_q$ and $\SPM_s$ that contain it and its distance to neighboring
vertices. For every degree one vertex $b_{pq}$, we store the cells of
$\SPM_p$ and $\SPM_q$ that contain it, which edge of $P$ it is on, and
if applicable its separation from neighboring degree one vertices on
the same edge. For each bisector, we store the data structure of
Theorem~\ref{thm:kds_bisector}. Our data structure
uses a total of $O(n+m)$ space.

It is not to difficult to see that this certificate structure captures
all external events. For collapse and expand events involving degree
three vertices we explicitly certify that the distance to its adjacent
vertices is non-zero. For events involving degree one vertices we
explicitly track which edge contains each such a vertex. This allows
us to detect vertex events. Furthermore, we maintain distance of each
degree one vertex to other degree one vertices on the same edge. Thus
we can detect 1,3-expand events. We also
maintain which cells of the $\SPM$ the vertex is contained in, which
allows us to detect 1,2-expand and $1,2$-collapse events. What remains
are the 2,2-events. These are detected by the data structure of
Theorem~\ref{thm:kds_bisector}.

\paragraph{Handling events} Handling the events is similar to what we
described in Sections~\ref{sub:A_KDS_for_maintaining_the_Bisector}
and~\ref{sub:A_Kinetic_Data_Structure_center}. Hence, we describe only
what is new or different here.

At all external-events we have to update the shortest path map
representations of the Voronoi cells. In most cases, this involves
adding or removing a single vertex to the shortest path map. This can
easily be handled using local computations in $O(\log^2 m)$ time.
The most expensive operation is computing the location of a new
Voronoi vertex, which may take $O(\log^2 m)$ time~\cite{oh_ahn2017voronoi_journal}.
In
case of vertex events, we may have to move an entire region in
$\SPM_s$ to $\SPM_p$. Since all shortest paths in such a region go via
the vertex involved, we can perform these updates in $O(\log m)$
time using the above data structure.

Since there are now $n$ sites, we maintain $O(n+m)$ certificates, and
thus updating the event queue takes $O(\log (n+m))$ time. Furthermore,
we now have multiple degree three vertices, and thus we have to handle
$3,3$-collapse and expand events. These are handled in a similar
fashion to the other events; we update the Voronoi regions, and
compute new certificates certifying the movement of the vertices
involved from scratch. All these updates can be done in
$O(\log^2 m+\log n)$ time.

At a vertex event where $p$ and $r$ are equidistant to a vertex $v$,
the region $R$ that moves from $\SPM_p$ to $\SPM_r$ may now be bounded
by a bisector $B_{rq}$ rather than $B_{pq}$ (see
Fig.~\ref{fig:bisector_split}). Since, at the time of the event, the
relevant parts of $B_{pq}$ and $B_{rq}$ coincide we can obtain the new
part of $B_{rq}$ by splitting $B_{pq}$, and updating the movement of
the associated sites. In particular, replacing the function expressing
the distance $p$ to $v$ by the distance from $r$ to $v$. Our bisector
KDS allows such updates in $O(\log^2 m)$ time (Theorem~\ref{thm:kds_bisector}).

Finally, we may have to update the certificates associated with the
Voronoi vertices as a result of changes to the individual shortest
path maps. For example, when a site $s$ can no longer see polygon
vertex $v$, this affects all Voronoi certificates of vertices for
which the shortest path goes through $v$. While our KDS for the
bisector (Theorem~\ref{thm:kds_bisector}) can update the affected
certificates of such a change efficiently, this unfortunately does not
hold for the certificates associated with degree one or degree three
vertices. Updating these requires $O(k(\log^2 m+\log n))$ time, where $k$ is
the number of neighbors of $s$ in $\VD_S(P)$. It is an interesting
open question to try and handle such events implicitly as well. We
therefore obtain the following result:

\begin{theorem}
  \label{thm:kds_voronoi_diagram}
  Let $S$ be a set of $n$ sites moving linearly inside a simple
  polygon $P$ with $m$ vertices. There is a KDS that maintains the
  geodesic Voronoi diagram $\VD_P(S)$ that uses $O(n+m)$ space and
  processes at most $O(m^3n^3\beta_z(n))$ events, each of which can be
  handled in $O(k(\log^2 m+\log n))$ time, where $k$ is the number of
  neighbors of the affected Voronoi cell.
\end{theorem}

\section{Refining the bounds}
For simplicity, the bounds in this paper have been expressed as a function of $n$ (the number of sites) and $m$ (the number of vertices of $P$). When dealing with simple polygons, it is common to use an additional parameter: the number of reflex vertices of $P$ (denoted by $r$). Since each reflex vertex is in particular a vertex, we immediately have $r \leq m$, but in general the two parameters are asymptotically different.

We believe that most $m$ terms that appear in our bounds can be turned into a function of $r$ instead. This can be achieved by looking at each of the bounds and carefully analyzing what depends on $m$ and what actually depends on $r$. Alternatively, we can use the polygon simplification technique of Aichholzer {\em et al.}~\cite{ahkpv-gpps-15}:

\begin{theorem}[Theorem 1 of~\cite{ahkpv-gpps-15}, rephrased]
Given a simple polygon $P$ of $m$ vertices and $r$ reflex vertices, there exists a simple polygon $P' \supseteq P$ of $O(r)$ vertices such that any shortest path in $P$ is a shortest path in $P'$.
\end{theorem}

This result is a general tool that can be used to transform many dependencies of $m$ into dependencies of $r$. Note that the focus of~\cite{ahkpv-gpps-15}  is algorithmic (i.e., they study at how fast $P'$ can be computed) whereas our proof is combinatorial. For our purposes, Aichholzer's result implies that  bisectors in $P$ must be contained in the bisectors of $P'$. Thus, we can track topological changes of bisectors in $P$ by focusing on $P'$ instead (and since $P'$ has $O(r)$ vertices, the bounds on the number of such events would depend on $r$).

This method can be used for most events but not all of them: for example, vertex
events happen when a bisector sweeps through any vertex of $P$ (and
thus the dependency on $m$ cannot be removed). Thus we wonder if there
is some technique similar to Aichholzer's that we could use to refine
all of our bounds and that we can obtain the correct dependency in $m$
and $r$. We leave this as an interesting open question for future work.

\section{Conclusion}
We constructed compact, responsive, local, and efficient kinetic data structures for maintaining bisectors and Voronoi centers of a geodesic Voronoi diagram. These data structures use linear space and process a worst-case optimal number of events. The bisector KDS handles each event in $O(\log m)$ time, while the Voronoi center needs $O(\log^2 m)$ time per event. Together, these data structures form a compact KDS for maintaining the full augmented geodesic Voronoi diagram. For each type of event we provide upper bounds of the worst-case number of events (which ranges from cubic to fifth power). These bounds are close to the truth, since for each case we provide an instance whose number of events is tight or almost tight to the corresponding theoretical upper bound.

Obvious follow up questions are: to design a more efficient KDS for
this problem, refine the bounds (for example by using the number of
reflex vertices $r$ of the polygon), or to close the gaps between the
worst-case upper and lower bounds. A more interesting variation of the
latter problem would be to find more realistic instances; our
constructions are rather degenerate in that the polygons contain
narrow corridors that all sites pass through at controlled speeds,
causing many changes to the diagram. It would be interesting to define
a notion of \emph{realistic} or \emph{natural} input and study the
number of events in such situations. Such realistic input assumptions
have proven to be useful in related problems, for example when
analyzing the complexity of the area visible to a point on a
terrain~\cite{berg02realis_input_model_geomet_algor}.

There are two other ways in which our research can be extended. First, in the type of movement: in this paper we focused on sites that move linearly, but more general patterns could be considered. For example, can one design a compact, responsive, local, and efficient KDS for more general algebraic motions of bounded degree $d$? If so, what are the worst-case number of events that could happen (as a function of $d$)?

Another way to extend this would be to consider more general domains. Instead of a simple polygon one could look into polygonal domains (i.e., polygons with holes). Although many results extend to this space, the main difference when holes are added is that we can have paths of different homotopy connecting the same two points. Thus, any KDS that maintains the Voronoi diagram of moving sites must be able to compute which of them is the shortest one and detect when a different one becomes shorter. The result is that the complexity of most algorithms increases significantly and in some cases it is known that running times grow from linear to polynomials of degree larger than 10~\cite{DBLP:journals/comgeo/BaeKO19}. As an intermediate step, one could consider adapting our results to domains with $1$, $2$, or more generally $O(1)$ holes.

\paragraph{Acknowledgements}
{\small
  We would like to thank Man-Kwun Chiu and Yago Diez for interesting discussions
  during the initial stage of this research. We would also like to thank the anonymous reviewers which helped us significantly improve the presentation of this paper and made it more accessible to a wider audience.
}

\bibliography{kinetic}

\end{document}

%% file: 33flip_quadratic_m_lowerbound.tex
\begin{lemma}
  \label{lem:33collapse_lowerbound_m2}
  There may be $\Omega(m^2n)$ $3,3$-collapse events.
\end{lemma}

\begin{figure}
\centering
\includegraphics{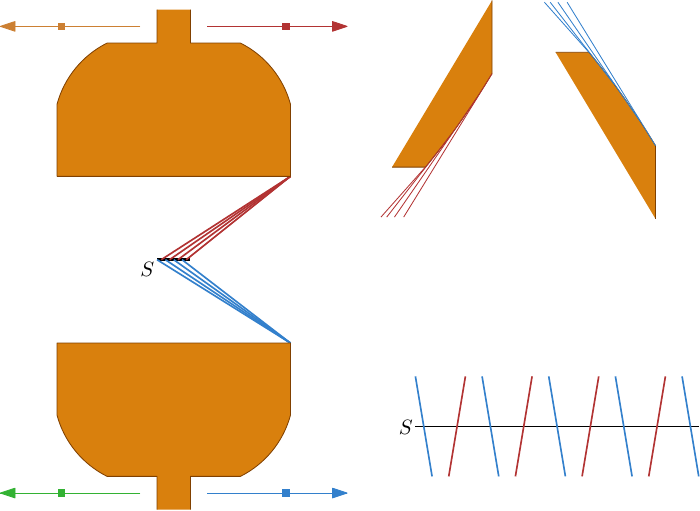}
\caption{An illustration of the lower bound construction for Lemma~\ref{lem:33collapse_lowerbound_m2}. The main construction is on the left, whereas the two figures on the top-right illustrated details of the right corners of the wine-glasses and the lower right is a vertically stretched depiction of the line segment $S$ and tangents with the convex chain on the corners.}
\label{fig:lowerbound_33collapse_m2}
\end{figure}

\begin{proof}
  The construction uses ideas similar to the ``wine-glass'' construction from Fig.~\ref{fig:lowerbound_22collapse_wineglass}. We will describe the construction in two steps as there are two different scale levels involved. The main construction is shown in Fig.~\ref{fig:lowerbound_33collapse_m2}, where we have two mirrored wine-glasses where the top of the wineglasses are right angles. If we assume the wineglasses and the four moving points are perfectly mirrored (we will add tiny deviations later) it follows that the four points are continuously co-circular with the centerpoint moving on a horizontal line in the middle between the two wine-glasses. By tailoring the slopes of the edges along the curved parts of the wine-glasses we can ensure that the centerpoint moves left to right along a horizontal line segment $\Omega(m)$ times. We denote this line segment by $S$.

   Next we add some variation to the two wineglasses. We replace the two right angled corners on the right with two convex chains with the following properties; (i) the lines aligned with the edges of the chain intersect the line segment $S$, (ii) the intersection points on $S$ alternate between lines aligned to edges of the upper chain and of the lower chain, and (iii) when moving from left to right along $S$ the nearest among the right two sites alternates. Note that for (iii) we will only consider the motion of the sites vertically above and below the wineglasses, so this statement does not depend on the exact location of the sites during the motion.

    The bound on the number of 3,3-collapse events can then be shown from these properties. First observe that the bisector between the two left sites is still a horizontal line and the portion of it that appears in the Voronoi diagram ends in a Voronoi vertex on $S$. As we can make the modification to the wineglasses arbitrarily small, the main motion of the bisector between the two upper (or the two lower) sites still remains the same. That is, it still sweeps from left to right along $S$. It follows that the Voronoi vertex that is the end of the bisectors of the two left sites also sweeps from left to right on $S$. By property (iii) the nearest among the two right sites alternates $\Omega(m)$ times, which means that they are equidistant $\Omega(m)$ times. So with each sweep of $S$, there are $\Omega(m)$ points where all four sites are equidistant and it follows that there are $\Omega(m^2)$ different 3,3-collapse events.

    We repeat the process with $\Omega(n)$ quadruples of points. Note
    that all events of a quadruple happen in some time interval. We
    make the wine glasses sufficiently wide so that these time
    intervals are disjoint, and so that the sites of quadruples that
    already have had their events are sufficiently far away from the
    center of the construction (segment $S$) that they do not
    interfere with later quadruples. Hence, each of the $\Omega(n)$ quadruples of points create $\Omega(m^2)$ events, resulting in the claimed bound. 
\end{proof}